# Data Science from 1963 to 2012

R. C. Alvarado
School of Data Science, University of Virginia
Last saved: 10/22/24 1:44:00 PM

**Abstract**: Consensus on the definition of data science remains low despite the widespread establishment of academic programs in the field and continued demand for data scientists in industry. Definitions range from rebranded statistics to data-driven science to the science of data to simply the application of machine learning to so-called big data to solve real world problems. Current efforts to trace the history of the field in order to clarify its definition, such as Donoho's "50 Years of Data Science" (Donoho 2017), tend to focus on a short period when a small group of statisticians adopted the term in an unsuccessful attempt to rebrand their field in the face of the overshadowing effects of computational statistics and data mining. Using textual evidence from primary sources, this essay traces the history of the term to the 1960s, when it was first used by the US Air Force in a surprisingly similar way to its current usage, to 2012, the year that Harvard Business Review published the enormously influential article "Data Scientist: The Sexiest Job of the 21ˢᵗ Century" (Davenport and Patil 2012), while the American Statistical Association acknowledged a profound "disconnect" between statistics and data science. Among the themes that emerge from this review are (1) the long-standing opposition between data analysts and data miners that continues to animate the field, (2) an established definition of the term as the practice of managing and processing scientific data that has been occluded by recent usage, and (3) the phenomenon of "data impedance"—the disproportion between surplus data, indexed by phrases like "data deluge" and "big data," and the limitations of computational machinery and methods to process them. This persistent condition appears to have motivated the use of the term and the field itself since its beginnings.

## 1. Introduction

Data science today is characterized by a paradox. The large number and rapid growth of job opportunities and academic programs associated with the field over the past decade suggest that it has matured into an established field with a recognizable body of knowledge. Yet consensus on the definition of data science remains low. Members and observers of the field possess widely variant understandings of data science, resulting in divergent expectations of the knowledge, skill sets, and abilities required by data scientists. Definitions—when they are not laundry lists—range from a rebranded and revised version of statistics to data-driven science to the science of data to simply the application of machine learning to so-called big data to solve real world problems. These differences cannot be reduced to mere "semantics"; they reflect deep-seated commitments to different values and understandings of knowledge, science, and the field's purpose. The lack of shared understanding poses a significant problem for academic programs in data science: it inhibits the development of standards and a professional commuity, confounds that allocation of resources, and threatens to undermine the authority and long-term prospects of these programs.

This essay presents a brief historical review of the use of the phrase "data science," and its grammatical variants "data sciences" and "data scientist," as a starting point for producing an authentic definition of the field as a coherent body of knowledge.[1] This history is presented as a series of milestones that delimit the periods in which the term takes on a new meaning. It is shown that these meanings are always additions to and inflections on existing meanings; in no case do they completely contradict what precedes them, nor do they appear as cases of random independent invention. The result is a picture of the transformation of a complex of meaning that indexes a set of technical, social, and cultural realities that have a continuous relationship to the current situation of data science, a situation that motivates the writing of this essay.

---

[1] Conway's famous Venn diagram comprising the areas of computer programming ("hacking"), math and statistics, and substantive expertise is no help here; each area is loosely defined, and the structure and purpose of their intersections are underspecified, a point underscored by the numerous reformulations of the diagram. This ambiguity may have been intentional, as Conway himself considered the term a "misnomer"(Conway 2010).

No claims are made for having discovered the first actual utterances of the term in question, neither at its origin nor during any of its transformations. Instead, the documentary record that comprises the sum of databases and documents available to the author, both online and off, is regarded as a kind of film, or perhaps an archaeological settlement pattern, on which collective verbal behavior impinges and leaves its marks. It is likely to be incomplete, but also comprehensive enough to capture patterns to a degree of resolution high enough to support the claims being made.

## 2. A Note on Method

The primary method employed is the close reading and precise seriation of textual evidence drawn from a representative collection of primary sources, including organizational reports, academic articles, news stories, and other contemporary forms of evidence. These are used to trace the history of the term's social and institutional contexts of use as well as its denotative and connotative meanings. Extensive extracts are often presented, rather then summarized, as these in many cases provide direct and illuminating evidence for the meanings in question.[2]

It is recognized that to trace the history of a term, must less a character string, is insufficient to represent the full history of what that term denotes, in this case a complex assemblage of concepts, tools, and practices that characterize data science today, many of which clearly precede or parallel the use of the term. Nevertheless, the exercise serves as a valuable starting point from which to develop a complete historical account, since finding textual examples for a term's string is relatively easy using textual databases, and because any related fields, such as data analysis or computational statistics, will be found to intersect with the term and may be pursed separately.

More important, although phrases like "data science" are, like all linguistic signs, arbitrary, they acquire motivation when they function as banners or brands under which allegiances are formed, catalyzing potential affiliations into actual ones. Such phrases are historically embedded speech acts with perlocutionary effects—they do not merely describe things in the world, they also instantiate them through their usage by agents, who influence the formation of their referents. This helps to explain why, once the phrase began to trend after 2008, many who previously would not affiliate themselves with the term began doing so, initiating a preferential attachment process to the term, and thereby complexifying its definition. It also explains the purpose of efforts to define the field, or to explain it away: each definition has a prescriptive dimension, since by proposing a "correct" definition, it attempts to influence usage and the field it denotes. The present essay is no exception.

Another reason for beginning with a history of words, via their traces as character strings, is that in historical research it is much easier to study words than the things they stand for, although

---

we often (conveniently) forget this relationship and conflate the two, believing we can easily move from language to the world. In our perceptions of the world beyond the ken of our immediate experience—and even there—we are enmeshed in language to a greater degree than we may like to admit. Words in the form of written documents (texts) constitute the primary source of data on which the construction of historical understanding depends. So, even though one may wish to get past words and study things as they are, the fact that these things are in the past, and mainly represented to us through documents and other material traces, means that one must begin with these. Ultimately, however, the purpose of working with written records is to get at the things they stand for and index, much as quantitative data $\mathcal{D}$ are used to construct an hypothesis $\theta$ to explain their existence. We may regard this phase of work as similar to the Bayesian task of establishing likelihoods and priors on the way to estimating posteriors.[3]

Finally, it is helpful to understand the larger theoretical lens through which these methods are applied and this history is presented and interpreted. The primary assumption is that science and all forms of knowledge production are cultural systems, in the sense proposed by Clifford Geertz and others (Geertz 2017; Martin 1998). This is not to affirm or deny the objectivity of science, or its effectiveness relative to other forms of knowledge production, but simply to assert that science, like all human endeavors, is made possible in and through social interaction over time and space and the media forms that make such interaction possible. Among these media forms the most significant is language; given this, I adopt a discourse-centered approach to understanding culture (Urban 1993). Futher, as a means to theorize the causal relationship between language and world, I adopt the view that discourse—spoken and written language, as opposed to generative grammar—is best uunderstood as situated action (Mills 1940; Suchman 1987; Norman 1993). In this view, language does not simply refer to the world, but participates in it, producing effects through its use in concrete situations. From an interpretive perspective, the meanings of words index the work they peform. This essentially causal conception of language use allows us to make sense of the relationship between terms like data science and the human endeavors with which they are associated, the posterior relationship encoded in the Bayesian framework described above.

## 3. Historical Sequence

### 3.1. 1963: Computational Data Science 1

The first recognizable uses of the phrase "data science" appear in the plural form in the early 1960s. Two main uses are found in the written record almost simultaneously, one in a military context, the other corporate.

---

[3] This is more than an analogy. If we think of the work of textual interpretation, on which historical research depends, in a probabilistic framework, we may express the hermeneutic relationship between words (*Sr*) and meanings (*Sd*) as follows:

$$P(Sd|Sr) = \frac{P(Sd)P(Sr|Sd)}{P(Sr)}$$

Inasmuch as *Sr* and *Sd* represent what Schleiermacher called the "linguistic" and "psychological" aspects of interpretation, the formula also expresses the logic of the hermeneutic circle as a matter of updating the prior and recomputing the posterior (Palmer 1969). The analogy between the Bayesian approach to causality and the hermeneutic approach to meaning has been noted by others (Groves 2018; Friston and Frith 2015; Ma 2015; Reason and Rutten 2018; Frank and Goodman 2012).

The military use appears in a series of reports covering the period from July 1962 to June 1970 on research carried out by the Data Sciences Laboratory (DSL), founded in 1963 as one of several labs associated with the US Air Force Cambridge Research Laboratories (AFCRL).[4] These reports do not provide an explicit definition of data science or a rationale for choosing the expression over others, but its meaning is clear from context. Consider the stated motivation for the lab—which, as the first attested use of the phrase, is worth quoting at length:

> The most striking common factor in the advances of the major technologies during the past fifteen years [i.e. since WWII] is the increased use and exchange of information. Modern data processing and computing machinery, together with improved communications, has made it possible to ask for, collect, process and use *astronomical amounts* of detailed data. …

> But in the face of this progress there is impatience with *the limitations of existing machines*. …

> A large number of military systems—for example, those concerned with surveillance and warning, command and control, or weather prediction—deal in *highly perishable information*. Few existing computers are capable of handling this information in "real-time"—that is, processing the data as they come in. Higher speed is one way to a solution. But increased speed will not overcome fundamental shortcomings of existing computers. These shortcomings arise from the fact that existing machines, having essentially evolved as numerical calculators, are not always optimally organized to perform the tasks they are called upon to do. …

> … *A considerable amount of the data to be processed is not numerical*. It is in audio or visual form. Immense amounts of visual data—for example, TIROS satellite pictures or bubble chamber pictures of atomic processes—remain unevaluated for lack of processing capability. In part this is due to the fact that, from the data processing point of view, the information content of pictorial inputs is highly redundant, *demanding excessive channel capacity* in transmission and compelling processing machinery to handle vast amounts of meaningless or non-essential information. Similar considerations prevail for speech. ...

> In real-life situations *data are almost never available in unadulterated form*, but are usually distorted or masked by spurious signals. Examples are seismic data, radio propagation measurements, radar and infrared surveillance data and bioelectric signals. …

> An increasing amount of *data processing research* is aimed at the creation of machines or machine programs that incorporate features of *deductive and inductive reasoning, learning, adaptation, hypothesis formation and recognition*. Such features are commonly associated with human thought processes and, when incorporated in machines, are frequently termed "artificial intelligence." Artificial intelligence is of utmost importance in decision situations where not all possible future events can be foreseen.(AFCRL 1963) (Emphases added.)

The two later reports are more succinct:

> The program of the Data Sciences Laboratory centers on the processing, transmission, display and use of information. Implicit in this program statement is an emphasis on computer technology (AFCRL 1967: 13).

> Broadly defined, the program of the Data Sciences Laboratory involves the automatic processing, filtering, interpretation and transmission of information (AFCRL 1970: 318).

Not expressed in these excerpts is the substantive research on pattern recognition and classification, machine learning, neural networks, and spoken language processing that constituted a significant amount of the lab's research portfolio.

---

[4] The DSL was formed by combining the Computer and Mathematical Sciences Laboratory and the Communications Sciences Laboratory in the 1963 reorganization (Venkateswaran 1963: 628). Within the AFCRL, the lab was noted for its "research on speech patterns dated back to the 1940's [sic]" (Altshuler 2013: 27-28).

Based on these excerpts alone, one could be forgiven for inferring that data science was invented by the US Air Force in 1963 with the formation of the DSL. Most of the elements currently considered central to the field were brought together there: a concern for processing what is later called "big data," clearly defined in terms of volume, velocity, and variety; a recognition of the fundamental messiness of data; and a focus on artificial intelligence as an essential approach to extract value from data.

Most important is the implicit meaning of the term and its historical motivation. The AFCRL was originally established in 1945 (under another name, the Cambridge Field Station) to keep the scientists and engineers who performed significant radar and electronics research in WWII. During the 1950s, the lab focused on the Lincoln Project, which led to the creation of the Semi-Automatic Ground Environment (SAGE). This was a system of large, networked computers that coordinated data from many radar sites and processed it to produce a single unified image of the airspace over a wide area.

Although responsibility for research on military surveillance systems was moved out of the lab in 1961, before the Data Sciences Lab was formed, it is plausible that the SAGE project influenced the mission of the lab by providing a concrete paradigm for a new kind of information processing situation. The was the situation of using of advanced computational machinery and state-of-the-art data reduction and pattern recognizing methods to process vast amounts of real-time signal data, coming from geographically distributed radars and satellites, in order to represent a complex space of operations and guide making decisions about how to operate in that space. The paradigm was applied to the problem of weather forecasting and other geophysical domains.

Evidence for the influence of wide-area, radar and satellite-based information systems on the conceptualization of data science may be found in the idioms we currently associate with the field, such as the use of "signal and noise" to refer to the presence and absence of statistically significant patterns and the use of Reciever Operator Characteristic (ROC) curves—first used by military radar operators in 1941—to measure the performance of binary classifiers. Other idioms, such as "data deluge" also emerge in this context. A history of this expression is worth its own study, but it is clear that its provenance is that of the situation described above. The term gained currency in the 1960s in reference to satellite data by NASA and the military. Consider this passage from the NASA publication *Scientific Satellites*:

> The data deluge, information flood, or whatever you choose to call it, is hard to measure in common terms. An Observatory-class satellite may spew out more than $10^{11}$ data words during its lifetime, the equivalent of several hundred thousand books. Data-rate projections, summed for all scientific satellites, prophesy hundreds of millions of words per day descending on Earth-based data processing centers. These data must be translated to a common language, or at least a language widely understood by computers (viz, PCM), then edited, cataloged, indexed, archived, and made available to the scientific community upon demand. Obviously, the vaunted information explosion is not only confined to technical reports alone, but also to the data from which they are written. In fact, the quantity of raw data generally exceeds the length of the resulting paper by many orders of magnitude (Corliss 1967: 157).[5]

---

[5] Preceding the usage of data deluge and in a wider context is "information explosion." Both expressions conjure images of disaster and have been remarkably persistent up to the present era. Only with the coining of "big data" have they been displaced by a more positive term.

Work on such projects generated an enormous amount of research on the problems of processing and interpreting data. In the preceding text, the author describes this work in some detail, specifying a series of stages in which data are transformed into a form suitable for scientific analysis. We would recognize this work today as data wrangling. It is reasonable to infer that the concept of data science emerged to designate this kind of work, which, in any case, is consistent with the published mission of the Data Sciences Lab. Indeed, prior to the formation of the DSL, the category of "data-processing scientist" was in use to designate the work involved in data reduction centers, such as the one built at the Langley Aeronautical Lab in Virginia to process the enormous amounts of data generated by wind tunnel experiments and other sources associated with the nascent space program. Data reduction was essential to projects like SAGE, in which vast amounts of real-time signal data had to be reduced prior to analysis. In a House appropriations hearing in 1958, the following description of this kind of work was provided by Dr. James H. Doolittle,[6] the last chairmain of NACA before it became NASA:

> The data processing function is much more complex than the mere production line job of translating raw data into usable form. Each new research project must be reviewed to determine how the data will be obtained, what type and volume of calculations are required, and what modifications must be made to the recording instruments and data-processing apparatus to meet the requirements. *It may even be necessary for the data-processing scientist to design and construct new equipment for a new type of problem.* Some projects cannot be undertaken until the specific means of obtaining and handling the data have been worked out. In some research areas, on-line service to a data processing center saves considerable time by allowing the project engineer to obtain a spot check on the computed results while the facility is in operation. This permits him to make an immediate change in the test conditions to obtain the results that he wants (Appropriations 1958: 147).

Here data science designates a kind of research focused on what we may call the *impedance* that arises from the ever-growing requirements of data, produced by an expanding array of signal generating technologies (e.g. satellites), scientific instruments, and reports on one hand, and the limited capacity of computational machinery to process them on the other. It is concerned specifically with the development of computational methods and tools to handle the problems and harness the opportunities posed by surplus data. In this sense, data science is the science of handling and extracting value from data by means of computation; it is focused on computational data processing. Although the specific technologies have changed continually, the condition of data impedance, the disproportion between data abundance and computational scarcity relative to the need to extract value from the data, has been constant since this time, and defines the condition that gives rise to data science in this sense.

This interpretation of the meaning of the term is corroborated by other contemporary usages. A report on a US Department of Defense program to define standards "to interchange data among independent data systems" refers to a "Data Science Task Group" established in 1966 "to formulate views of data and definitions of data terms that would meet the needs of the program" (Crawford, Jr. 1974: 51). Crawford, a fellow student of Claude Shannon at MIT under Vannevar Bush, was affiliated with IBM's Advanced Systems Development Division, a group that had developed optical scanners to recognize handwritten numbers in 1964. In addition, the term appeared in the trademarked name of at least two corporations in the United States: Data Science Corporation, formed in 1962 by a former IBM employee (*St. Louis Post-Dispatch* 2014), and Mohawk Data Sciences, founded in 1964 by a three former UNIVAC engineers (*The New York Times* 1966). Both companies provided data processing services and lasted well into the era of personal computing. In the late 1960s and 1970s, many other companies used term as well, such as Data Science Ventures (Mort Collins Ventures n.d.) and Carroll Data Science Corporation (Office 1979).[7]

---

[6] This is the very same General Doolittle of Doolittle's Raid.

Here we should pause to consider the meaning and significance of the word "data" in these examples, especially given the DOD's concern to define it, as a clue for the motivation of the term "data science" when other candidates, such as computer science and information science, might have sufficed at the time. The choice of the term appears to be motivated by a concern to define and understand *data* itself as an object of study, a surprisingly opaque concept that is thrown into sharp relief in the context of getting computers to do the hard work of processing information in the context of impedance, as a result of their commercialization and widespread use in science, industry, and government. Thus although the term "data" has a long history, deriving from the Latin word for that which is *given* in the epistemological sense, either through the senses, reason, or authority, in this context it refers to the structured and discrete representation of information sources so that these may be processed by computers. In other words, *data is machine readable information.*[8] It follows that the data sciences in this period are concerned with understanding machine readable information, in terms of how to represent it and how to process it in order to extract value.

Further evidence of this concern for what might be called the information crisis in scientific research—and for the idea that the solution to this crisis hinges on refining the concept of data—can be found in the formation of the International Council for Science (ICSU) Committee on Data for Science and Technology (CODATA) in 1966. This organization was established by an international group of physicists alarmed that the "deluge of data was swamping the traditional publication and retrieval mechanisms," and that this posed "a danger that much of it would be lost to future generations" (Lide and Wood 2012). Importantly, CODATA still exists and currently identifies itself with the field of data science. In 2001 it launched the *Data Science Journal,* focused on "the management, dissemination, use and reuse of research data and databases across all research domains, including science, technology, the humanities and the arts" (*Data Science Journal* n.d.). Aware that the definition of the field had changed significantly since its founding, the journal provided the following clarification in 2014:

> We primarily want to *specify* our definition of "data science" as the classic sense of the science of data practices that advance human understanding and knowledge—the evidence-based study of the socio-technical developments and transformations that affect science policy; the conduct and methods of research; and the data systems, standards, and infrastructure that are integral to research.
>
> We recognize the contemporary emphasis on data science, which is more concerned with data analytics, statistics, and inference. We embrace this new definition but seek papers that focus specifically on the data concerns of an application in analytics, machine learning, cybersecurity or what have you. We continue to seek papers addressing data stewardship, representation, re-use, policy, education etc.
>
> Most importantly, we seek broad lessons on the science of data. Contributors should generalize the significance of their contribution and demonstrate how their work has wide significance or application beyond their core study (Parsons 2019; emphasis in original).

---

[7] This continues into the 1980s, with Gateway Data Sciences Corp and Vertex Data Science, Ltd.

[8] Throughout this essay, the word "data" in the singular is intented to signify the abstract concept of data. When used in the plural, it refers to an concrete set of data. A datum refers to a concrete element of such a set.

This retrospective definition supports the idea that data science in the 1960s—which we may call, following this note, classical data science—was concerned with understanding data practices, where data is understood to be a universal medium into which information in a variety of native forms, from scientific essays to radio signals from outer space, must be encoded so that it may be shared and processed. Data science as "the science of data practices that advance human understanding and knowledge" is concerned with defining and inventing this medium, it's structure and function.

Tukey's famous essay on data analysis, which appears during the same time period, touches on some of the drivers noted here, such as the high volume and spottiness of real data and the impact of the computer, but from the perspective of advanced mathematical statistics (Tukey 1962). One difference between his view and that adopted by the AFCRL is of interest here: whereas Tukey appears to have regarded the computer as a more or less fixed technology, replaceable in many tasks by "pen, paper, and slide rule" but irreplaceable (he conceded) in others, the Data Sciences Lab viewed the computer as a fluid technology, one that needed to be pushed beyond its original design envelope as a numerical calculator. In fact, the AFCRL and similar groups appear to have provided the impetus to move computer science beyond a concern for abstract algorithms and to include the study of data structures and technologies, specifically databases. It is, as we shall see, a difference that continues to underlie current disputes over the meaning and value of data science.

## 3.2. 1974: Computational Data Science 2

It is clear that by the early 1970s the term data science had been in circulation in several contexts and referred to ideas and tools relating to computational data processing. Importantly, these usages were not obscure—the AFCRL was one of the premier research laboratories in the world and closely connected with Harvard and MIT (Altshuler 2013), an international cross-roads of intellectual life where many would have come into contact with the term. Similarly, IBM and UNIVAC, the sources of the founders of two self-proclaimed data science companies, were the two largest computer manufacturers at the time.[9]

Although the AFCRL closed the Data Sciences Lab in 1972, the term continued to be used, most notably by the Danish computer scientist Peter Naur, who suggested that computer science, a relatively new field, be renamed to data science. His argument, consistent with previous usage, was that computer science is fundamentally concerned with data processing and not mere computation, i.e. what the AFCRL derided as numerical calculation. Earlier, in the 1960s, Naur had coined the term "datalogy" (Danish: *datalogi*) for this purpose, but later found the term data science to be a suitable synonym, perhaps due to its currency or to his familiarity with the DSL, which shared his research interest in developing programming languages (Naur 1966). In

---

[9] As further evidence of the visibility of the DSL at the AFCRL to the wider world, consider that the following advertisement that appeared in a 1964 issue of the British weekly *Nature*:

A SYMPOSIUM on "Models for the Perception of Speech and Visual Form", sponsored by the Data Sciences Laboratory of the Air Force Cambridge Research Laboratory, will be held in Boston during November 11-14. Further information can be obtained from Mr. G. A. Cushman, Wentworth Institute, 550 Huntington Avenue, Boston, Massachusetts 02115 ("Announcements" 1964) ("Announcements" 1964).

contrast to the AFCRL, Naur provided an explicit definition of data science, which he summarizes on his website:

> The starting point is the concept of *data*, as defined in [Gould, 1971]: DATA: *A representation of facts or ideas in a formalized manner capable of being communicated or manipulated by some process.* Data science is the science of dealing with data, once they have been established, while the relation of data to what they represent is delegated to other fields and sciences.

> The usefulness of data and data processes derives from their application in building and handling models of reality.

> …

> A basic principle of data science is this: The data representation must be chosen with due regard to the transformation to be achieved and the data processing tools available. This stresses the importance of concern for the characteristics of the data processing tools.

> Limits on what may be achieved by data processing may arise both from the difficulty of establishing data that represent a field of interest in a relevant manner, and from the difficulty of formulating the data processing needed. Some of the difficulty of understanding these limits is caused by the ease with which certain data processing tasks are performed by humans (Naur n.d.; emphasis and citation in original).

Clearly, Naur's definition inherits the classical definition described above; it locates the meaning of the term in the series of practices associated with the larger activity of data processing. These practices include establishment, choice of representation, conversion and transformation, the modeling of reality, and the guiding of human actions.

One difference is that Naur is keen to locate data science within a division of labor implied by this general process, separating data science *per se* from the work of data acquisition (establishment) and the domain knowledge required to acquire data effectively. In this view, data science is more specifically concerned with the formal representation of data (i.e. with data structures and models), a practice that must be done in light of how data are to be transformed downstream, and with which tools (i.e. algorithms and programming languages). As we shall see, the weighting that Naur assigns to this kind of work is not inherited by later theorists. However, the general image of a sequential process with distinct phases in the life cycle of data is. Here we see the appearance of the image of a pipeline, unnamed but implied by the concept of *process*, which dominates the mental representation of the field from its origins in the 1960s.

Far from being a fluke, Naur's usage developed the classical definition of data science initiated by NASA and the Air Force, intentionally or not. The fact that his attempt to rename computer science failed outside of his native country (and Sweden) is not important; his understanding of computer science sheds light on how closely the concept of data was (and is) related to computation and process.

It is worth noting that Naur's definition implies a familiarity with the real-world provenance of data processing in industry and government. Indeed, by this time computational data processing had penetrated all sectors of society, and the pressure to improve tools and methods to represent and process data had increased as well. As a result of this pressure, two important data standards were developed in this timeframe: Codd's relational model, which laid the foundation for SQL and commercially viable relational databases in the 1980s, and Goldfarb's SGML, which would become a standard for encoding unstructured textual data (such as legal documents) and later the basis for HTML and XML (Codd 1970; Goldfarb 1970). This focus on the human context of data processing is reflected in his later work; a volume of selections of his writing from 1951 to 1990, which includes his essay on data science, is entitled "Computing: A Human Activity" (Naur 1992).

After Naur, the term receded into the long tail of usage. Companies continued to use it (mostly in the plural), and it also appeared in the name of a unit within the US National Oceanic and Atmospheric Administration (NOAA), Environmental Data Science (*Library Journal* 1977), tasked with managing a growing collection of environmental data sets. For its part, the term datalogy continues to be used in Denmark and Sweden.

Interestingly, in 1977 a prefixed variant of the term does appear in the title of the technical report, "Non-Parametric Statistical Data Science: A Unified Approach Based on Density Estimation and Testing for 'White Noise'" (Parzen 1977). However, Parzen later publishes a version of this work as "Nonparametric Statistical Data Modeling" (Parzen 1979), indicating that his original word choice was mistaken. Yet the original choice may not have been entirely unmotivated: Parzen's work attempts to unify parametric and non-parametric methods under one umbrella. Given the natural inclination of mathematical statistics for the former and data analysis for the latter, his choice of the term data science may have signaled an attempt to encompass both approaches to data. It is also worth noting that he later used to the term to introduce a "new culture of statistical science called LP MIXED DATA SCIENCE" (Parzen and Mukhopadhyay 2013), after the term became popular.[10] Whether or not this unifying goal was his motivation, statisticians later became quite interested in the term for precisely this reason.

### 3.3. 1992: Statistical Data Science 1

In the early 1990s, the term resurfaced in a context that proved to be more enduring. It appeared in the title of a 1994 essay by the Japanese statistician Noburu Ohsumi on the application of hypermedia to the problem of organizing data, "New Data and New Tools: A Hypermedia Environment for Navigating Statistical Knowledge in Data Science," an elaboration of an essay published two years earlier (Ohsumi 1992; 1994).[11] In these essays, Ohsumi described the by now familiar litany of problems associated with data impedance, although this time the focus was on the production of data resulting from its analysis and storage, not its consumption in so-called raw form:

> In research organizations handling statistical information, the volume of stored information resources, including research results, materials, and software, is increasing to the point that conventional separate databases and information management systems have become insufficient to deal with the amount. Increasing diversification in the media used these days interferes with the rapid retrieval and use of the information needed by users. A new system that realizes a presentation environment based on new concepts is needed to inform potential users of the value and effectiveness of using the vast amount of diverse data (Ohsumi 1992: 375).

For research facilities around the world, the product of classical data science—the database and data processing software—had become a sorcerer's apprentice, creating new problems with each solution. Organizations were drowning in the data sets they produced or acquired, the software used to process them, the print and digital libraries of reports and articles resulting from their analyses, and a host of other materials. The requirements, approach, and design goals of Ohsumi's proposed system, the Meta-Stat Navigator, are strikingly similar to those of a contemporary system designed to solve the information problems of another scientific organization: Berners-Lee's World Wide Web, famously developed at CERN in 1989 (Berners-Lee and Fischetti 2008). Of course, the latter quickly obviated Ohsumi's proposal and become synonymous with the Internet, invented decades earlier.

---

[10] Parzen's use of the term "culture" here echoes his comments on Breiman's famous essay on two cultures of statistical models, where he suggested that there are in fact several cultures, including his own, to which he devoted the majority of his response (Breiman 2001: 224–226).

[11] According to Ohsumi, "the term 'data science' appeared for the first time" in 1992, at a research exchange meeting between French and Japanese data analysts (so-called) at Montpellier University II in France (Ohsumi 2000: 331). He also claims to have "argued the urgency of the need to grasp the concept 'data science'" in 1992 (Ohsumi 2000: 329).

The significance of Meta-Stat for our purposes is that this kind of work was understood clearly as data science at this point in history. Data science continued to be connected with the processing and representation of data, and was distinct from data analysis, but with this important development: statisticians had become embedded in these technologies, and their work had changed significantly as a result. And, as a result of this change in working conditions, the connection between data analysis and data science became closer.

Here we may locate with some precision a crucial transformation in the meaning of the term, associated with its adoption by a new set of users. One clue to this change is the opportunity Ohsumi observed amidst the challenges posed by data deluge:

> … the information handled by the statistical sciences lies on the boundaries of various other sciences and clarifies the relationships and nature of information that joins these sciences. Development of a system that fully organizes and integrates strategic information is essential (Ohsumi 1992: 375).

The Meta-Stat "system," which we may take as a stand-in for data science itself, was designed to realize the opportunity opened up by the central position statisticians had come to occupy among the prolifically data-generating sciences and the computational environment in which these data were made available. Data science, in this view, is *meta-statistics*, an encompassing concern for understanding data, understood as a universal medium, and its relationship to knowledge. This perspective would be adopted by Ohsumi's senior compatriot and fellow statistician, Chikio Hayashi, whom Ohsumi described as "the pioneer and founder of data science" (Ohsumi 2004: 1).

In 1993, at a roundtable discussion during the fourth conference of the International Federation of Classification Societies held in Paris (IFCS-93), Hayashi uttered the phrase "Data Science" and was then asked to explain it. At the next conference (IFCS-96), he presented an answer, in addition to having the conference named to emphasize the importance of the term—"Data Science, Classification, and Related Methods." His definition is as follows:

> Data science is not only a synthetic concept to unify [mathematical] statistics, data analysis and their related methods but also comprises their results. It includes three phases, design for data, collection of data, and analysis on data. Data science intends to analyze and understand actual phenomena with "data." In other words, the aim of data science is to reveal the features of the hidden structure of complicated natural, human and social phenomena with data from a different point of view from the established or traditional theory and method. This point of view implies multidimensional, dynamic and flexible ways of thinking (Hayashi 1998: 41).

Hayashi goes on to describe the sequence of design, collection, and analysis as a primary and iterative "structure finding" process in which data are transformed from a state of "diversification," given the inherent "multifariousness" of the phenomena they represent, to one of "conceptualization or simplification" (41). The discovery of structure is accomplished with what we would recognize today as the methods of exploratory data analysis and unsupervised learning. In effect, Hayashi's definition abstracts the design goals of Ohsumi's Meta-Stat and presents them as "a new paradigm" of science, one that would encompass statistics, data analysis, and their vast output of data within in a unified, process-oriented framework—data science (40).

In addition to Hayashi's own definition, it helpful also to see how the field was defined by the editors (who included Hayashi) of the proceedings of IFCS-96:

> The volume covers a wide range of topics and perspectives in *the growing field of data science*, including theoretical and methodological advances in domains relating to data gathering, classification and clustering, exploratory and multivariate data analysis, and knowledge discovery and seeking.

> It gives a broad view of the state of the art and is intended for those in the scientific community who either develop new data analysis methods or gather data and use search tools for analyzing and interpreting large and complex data sets. Presenting a wide field of applications, this book is of interest not only to data analysts, mathematicians, and statisticians but also to scientists from many areas and disciplines concerned with complex data: medicine, biology, space science, geoscience, environmental science, information science, image and pattern analysis, economics, statistics, social sciences, psychology, cognitive science, behavioral science, marketing and survey research, data mining, and knowledge organization (Hayashi 1998a: v; emphasis added).

Of interest here is use of "data science" as a big tent, an inclusive rubric under which to include a series of domains (which match roughly to a process) as well as a broad range of disciplines and levels, from tool builders to scientists and practice to theory. This passage is also significant for including within the scope of data science the methods of machine learning as well as data mining among the list of sciences concerned with "complex data," suggesting the prominence of these approaches at that time. We will see that not all definitions proceding from this community were as inclusive.

Hayashi assigns a revolutionary and almost messianic role to data science here. In his vision, the statistical sciences had lost their way. Mathematical statisticians had come to overvalue abstract inference and precision, and by choosing to work with the artificial data required to pursue these goals were "prone to be removed from reality" (40). Data analysts, although working with real data, had "come to manipulate or handle only existing data without taking into consideration both the quality of data and the meaning of data … to make efforts only for the refinement of convenient and serviceable computer software and to imitate popular ideas of mathematical statistics without considering the essential meaning" (40). As a result of these divergent attitudes toward data, and the disregard of both for the scientist's engagement with the primary, existential relationship between data and phenomena, the field had become stagnant and lacking in innovation. Data science emerges as a savior, unifying a divided people, showing their way out of the wilderness, and restoring prosperity and prestige to their community.

If Hayashi's criticisms of data analysis sound familiar to those leveled today against data scientists, it is because the issues data science was meant to resolve are recurring and systemic. So too is the separation between data analysis and mathematical statistics, which was recognized by Box, and later Tukey, in the 1970s. In his response to Parzen—who, we noted, sought to overcome a methodological split between the two subfields—Tukey wrote:

> I concur with the general sentiments expressed by George Box in his Presidential Address … that we have great need for the whole statistician in one body—for the analyst of data as well as for the probability model maker—and the inferential theorist/practitioner. One cannot, however, make a whole man by claiming that one can subsume one important class of mental activity under another class whose style and purposes are not only different but incompatible. To be "whole statisticians" as Box might put it, or to be "whole statistician-data analysts" as I might, means to be single persons who

> can take quite different views and adopt quite different styles as the needs change. As the title of my paper of yesterday put it, "we need both exploratory and confirmatory"! The twain can—and should—meet, but they need to remain a pair (or two distinct parts of a larger team) if they are to do what they should and can (Tukey 1979: 122).

Tukey implies a solution to the schism, later observed by Hayashi, in better organization, not in a utopian "new man" or in a synthetic science *per se*, recalling the division of labor proposed by Naur, but here focusing on different roles within that division. Implicit in this approach is the view that the problem with statistics was not epistemic but organizational.

Here it is helpful to recall a property of Kuhn's concept of paradigm—an obvious lens through which to observe our topic—which is often overlooked by those who use the term: it refers no to an abstract body of ideas that succeed on the basis of their intrinsic rationality or truth value, but to the successful practical application of ideas by means of novel methods and tools in a way that they may be imitated. The concept has both epistemic and social dimensions. Viewed in this light, the question of whether data science is in fact a science—our main question—becomes a matter of determining whether it solves important problems in new ways, by means of an assemblage of ideas, methods, and tools that may be grasped and imitated by others. Hence, although Tukey and Hayashi may appear to be divergent in their approaches to overcoming the problems, they represent the two aspects of a scientific paradigm, the one conceptual, the other practical. This should not be viewed as contradictory.

Following the IFCS meetings, as well as two meetings of the Japan Statistical Society that held "special sessions on data science," Ohsumi developed Hayashi's definition as well as its rationale (Ohsumi 2000: 331).[12] In a paper that explicitly addressed the relationship between data analysis and data science, and which is perhaps the first of several to claim the flag of data science for statistics, Ohsumi declared that because of its privileging of "mathematical methodologies" over an engagement with data acquisition, data analysis had become "a canary that has forgotten to sing," referring to a Japanese children's song that contemplates a silent bird's fate (332).[13] Amplifying Hayashi, he asserted that "[h]ow data are gathered is the key to defining the relevant information and making it easy to understand and analyze" (331). In making this point, Ohsumi referred to a new figure on the scene, one that contradicted the principles he proposed:

> In my opinion, this viewpoint on the meaning of data science is fundamentally different from data mining (DM) and knowledge discovery (KD). These concepts are not of practical use because they neglect the problems of 'data acquisition' and its practice (332).

It is significant that Ohsumi excludes these new fields—or field, since the two so frequently co-occur, along with the variant KDD, "knowledge discovery in databases"—from his definition of data science, since many today would consider the two synonymous. The paradox is instructive: the name "data mining," as used here, makes its appearance in the early 1990s as a rubric that included a set of practices motivated by precisely the same conditions that led the Tokyo school to propose the field of data science in the first place. Among these conditions was the relatively sudden appearance of vast amounts of data stored in databases—one of the fruits of classical data science—owing to the success of relational databases and personal computing in the 1980s, and a suite of tools to work with data, from spreadsheets to programming languages to statistical software packages. Whereas many statisticians viewed these developments with alarm, being acutely aware of the epistemic disruptions they produced for the received workflow of data analysis, the data mining community embraced them as an opportunity to convert data into value. Coming mainly from the field of computer science, data miners developed a set of methods that included the application of machine learning algorithms to the data found in databases in various contexts, from science to industry (such as point-of-sale records generated as by-product of computerized cash registers and credit card use). The relationship between machine learning and data mining was also mutually beneficial—data miners supplied machine learning with the large sets of data required for this class of algorithms to perform well. This relationship was greatly reinforced with the rise and development of the Web and social media platforms, which generated enormous amounts of behavioral data.

---

[12] At this point we shall call this the Tokyo school of data science, given the association of both Hayashi and Ohsumi with the Institute of Statistical Mathematics in Tokyo, Japan.

[13] The specific reference is to a poem, later set to music, written by the Japanese poet Saijoo Yaso (西条八十), who lived from 1892 to 1970. According to Miriam Davis, "The moral of the song is that if the canary loses its song it is not worth its existence so it should make the most of the gift of song it has been given." (Davis, n.d.)

Although the two fields—for simplicity, let's call them data analysis and data mining—were responding to the same conditions of data surplus and impedance, their philosophical orientations could not have been more opposed. This difference is clearest in their respective evaluations of data *provenance*, the source and conditions under which data are produced. For the data miner, data provenance is largely irrelevant to the possibility of converting data into value. Data are data, regardless of how they are generated, and the same methods may be applied to them regardless of source, so long as their structure is understood (e.g. time series). (Indeed, for the data miner data exists much as natural resources do, as a given part of the environment, which helps explain the success of the metaphor of *mining* over competing variants, such as *harvesting*, which implies intentional creation.) For the data analyst, as Hayashi and Ohsumi took such pains to emphasize, provenance is, or should be, everything, echoing the statistician's orthodox preference for experimental over observational data.

In defining data science in opposition to data mining, Ohsumi explains:

> Owing to the qualitative and quantitative changes in data [produced by the conditions described above], it is, indeed, becoming increasingly difficult to grasp all aspects of a dataset in explaining various phenomena. Therefore, new techniques, such as DM, KD, complexity, and neural networks, are being proposed. However, the potential of these methods to solve any of these problems is questionable (332).

Ohsumi goes on to characterize the way data has changed by listing the new kinds of data with which the statistician is confronted. These include prominently data sets found in databases as by-products of various processes, such as passive accumulation (e.g. from point-of-sale devices), unstructured data (included in text fields), and aggregated data generated "spontaneously and accumulating automatically in the electronic data collection environment" (332-333). He explained his concern with data mining:

> When it comes to analyzing these datasets, people discuss DM and related techniques. However, the important questions to answer are: what dataset is necessary to explicate a certain phenomenon, why is it necessary, how to design its acquisition, and how difficult the whole process is. *This is more important than the dataset itself*. Books on DM do contain terms such as "data preparation", "getting the data", "sampling procedures", and "data auditing", but there is an assumption that the dataset is given and the procedure may start with analysis. Fiddling with a dataset once it is collected is merely a self-contained play of data handling (Ohsumi 2000: 333: emphasis added).

Although his evaluation of data mining seems to be woefully off base—a great deal of Google's success, to take one example, was founded on their embrace of data mining at the time of Ohsumi's essay—in fact his concern is not with the success of predictive analytics *per se*, but with solving what he considered to be the central problem of data science, that of understanding how data are generated in the first place. Given some of the issues that classifiers have encountered with respect to racial bias, for example, he cannot be said to have been wrong.

Perhaps the most eloquent and authoritative account of the difference between data analysis and data mining is found in Leo Breiman's contemporary essay on an analogous pairing, what he called, echoing C. P. Snow, the "two cultures" of statistical modeling (Breiman 2001). In brief, one culture seeks to represent causality—the black box of nature that generates the empirical data with which statistics begins—by means of probabilistic or stochastic data models. The parameters, random variables, and relationships that compose these models are imagined to correspond to things in the world, at least in principle. Data are used to estimate the parameters of these models. This is the "data modeling culture," associated with traditional statistics and data analysis. Breiman guessed this culture comprised 98% of all statisticians, broadly conceived. The other culture bypasses attempts to directly model the contents of the black box and instead focuses on accounting for the data by means of goal-oriented algorithms, regardless of the correspondence of these to the world. This is the "algorithmic modeling culture," associated with computer science, machine learning, and, we might add, data mining. Breiman described the growth of this culture as "rapid" (beginning circa 1985) and characterized its results as "startling" (Breiman 2001: 200).

As Ohsumi wrote, for one culture the data models are more important than the data, and not all data are suitable to supporting the development of good data models. Hence the emphasis on design for data—the most important phase of data science is in the careful acquisition of data. For the other, data are both abundant and intrinsically valuable, and to a great extent have the power to account for themselves. Whereas the former is highly selective about the data it employs, and views with great suspicion—as we have seen—new forms of data coming from databases in a variety of formats, the latter embrace these data, and are not daunted by their size and complexity. On the contrary, these qualities are essential to the methods applied.

The point of Breiman's essay was to convince the 98% that their commitment to correspondence models had led to "irrelevant theory and questionable scientific conclusions" about underlying mechanisms. Perhaps more important, he argued that their priestly avoidance of impure algorithmic methods and data "not suitable for analysis by data models" (i.e. the accidental data found in databases, as opposed to data created by design) had prevented "statisticians from working on exciting new problems" (199–200). The canary had forgotten to sing, but for reasons precisely opposite to that claimed by the Tokyo school, whom Breiman may have admonished for an excessive concern for the conditions of acquisition.

One way to account for the difference between the two cultures is to look at their institutional settings. The data modeling culture is closely aligned with the project of academic science and the search for intelligible models of nature, whereas the latter are more associated with business needs, the pragmatic decision-making requirements of those clients who own the databases in the first place. Parzen, in his comment to Breiman's essay, characterizes this as the difference between "science" and "management" (224). This difference is reflected in Breiman's own biography, which is that of a liminal figure in this binary. He spent significant amounts of time as both an "academic probabilist" and as a free-lance consultant to industry and government, where he "became a member of the small second culture."

These different value orientations—deriving from the purpose for which one works with data in the first place—are reflected in their attitudes toward data and models. For one group, models are the capital on which one builds a career and a name. One wins a Nobel Prize for a successful model of the world, not for collecting the data upon which it was built, which are often forgotten and poorly documented. In business, however, models come and go, but the data constitute an irreplaceable form of capital, often taking years to accumulate and jealously guarded. Thus, for one group, models precede data; for the other, data precedes models. We might characterize the former as *essentialist* and the latter as *existentialist*, given the analogy that data : models :: existence : essence.

Breiman's essay marks a significant shift in the history of data science, a reversal in how data are regarded in relation to models. Consider that the phrase "data mining" itself, which was actually used by econometric statisticians to refer to the frowned upon practice of fishing for models in the data, of letting data specify models, a usage dating back to 1966 and at least up to 1995 (Lovell 1983; Ando and Kaufman 1966; Hendry 1995: 544). In his review of the concept in 1983, Lovell's remarks make it clear that the two usages are not entirely unrelated:

> The development of data banks ... has increased tremendously the efficiency with which the investigator marshals evidence. The art of fishing over alternative models has been partially automated with stepwise regression programs. While such advances have made it easier to find high $\overline{R}$'s and "significant" $t$-coefficients, it is by no means obvious that reductions in the costs of data mining have been matched by a proportional increase in our knowledge of how the economy actually works (Lovell 1983: 1).

> When a data miner uncovers $t$-statistics that appear significant at the 0.05 level by running a large number of alternative regressions on the same body of data, the probability of a Type I error of rejecting the null hypothesis when it is true is much greater than the claimed 5% (Lovell 1983: 1).

> It is ironic that the data mining procedure that is most likely to produce regression results that appear impressive in terms of the customary criteria is also likely to be the most misleading in terms of what it asserts about the underlying process generating the data under study (Lovell 1983: 10).

The fact that the same phrase, with a common referent but opposite sentiments, would be used contemporaneously is an indication of the social distance between the two cultures. But also, we can see that the approach Breiman proposed was exactly what was criticized by these statisticians: among the perceptions, or principles, he acquired as a consultant to work successfully with data, he specified the "[s]earch for a model that gives a good solution, either algorithmic or data" (201), a definition of data mining that would fit among those quoted, with some humor, by Lovell. In fact, the meaning of data mining, even among statisticians, changes during this period, going from a bad habit to a hot new area of research. Its negative evaluation by some, however, has persisted.

In defense of data miners against the criticisms of econometric statisticians and those of the Tokyo school, their focus on already collected data reflects Naur's view that data science should focus on representation and transformation, and not on establishment and domain knowledge—precisely the areas on which the Tokyo school focused. But more important, data miners had discovered something that data analysis had not, at least not as a shared perspective: data in fact do have a certain autonomy with respect to their provenance, and a variety of methods, including many from statistics, were revealing an entirely new and quite radical paradigm of science—one without need of "theory" (Anderson 2008a). "Self-contained play" actually pays off. In a certain sense, data miners were carrying out a principle asserted by Claude Shannon in his groundbreaking essay on information theory—that the "semantic aspects of communication are irrelevant to the engineering problem" (Shannon 1948: 5). The engineering problem in this case being the ability to discover significant patterns among features and to make predictions, and the semantics being the relationship between phenomena and data.

At the most general level, the epistemological orientations of the two cultures can be described by reference to how each understands the proper relationship between data and models on the one hand and motivating questions on the other. For the traditional data analyst, one acquires data and develops models in order to answer scientific questions. These derive from established fields ranging across the natural, life, and social sciences, from which there is no shortage of compelling problems to solve. For the data miner, the relationship is reversed: the presence of abundant data, found in databases, creates a need to find value in them, a vacuum to fill. Although, as we have seen, the field is sometimes called knowledge discovery, it might better have been called *question* discovery. Consider this sentence, drawn from an early essay on KDD: "American Airlines is looking for patterns in its frequent flyer databases" (Piatetsky-Shapiro 1991). This is not something a data analyst would utter publicly.

It is hard to overestimate the width of the gap between these two fields. To this day, statisticians, who view themselves as the inheritors and guardians of the scientific method, regard the unidirectional relationship between understanding why and how one collects data, and the collected data themselves, nearly as strongly as geneticists regard the relationship between genotype and phenotype—the arrow of information moves in one direction. Epigenetics notwithstanding, violation of this dogma is tantamount to heresy. The data miner has no such concern; data are data and data have value. The trick is to discover that value before anyone else does. This is not to say that data miners do not have questions in hand before working with data. Often clients have very specific questions, and existing databases are found that more or less match the requirements of the question. Indeed, in defending himself against Cox's charge of putting data before questions, Breiman wrote:

> I have never worked on project that has started with "Here is a lot of data; let's look at it and see if we can get some ideas on how we can use it." The data has been put together and analyzed starting with an objective (226).

But in these cases, the data miner is much more likely to work happily with these data and not wait for experimental data to be produced. If she does not succeed, she is as likely to blame her methods more than the data themselves. It is telling that in recounting his failure to come up with a predictive model of smog formation in Los Angeles, Breiman wished he had had "the tools available today," not better data (201).

## 3.4. 1997: Statistical Data Science 2

In his remarks to the commenters on his essay, among whom are Cox and Parzen, Breiman lamented: "Many of the best statisticians I have talked to over the past years have serious concerns about the viability of statistics as a field" (231). In addition to the evidence of the Tokyo school, this observation is corroborated by a series of papers and presentations produced by academic statisticians in the U.S. beginning in the mid-1990s, all of whom expressed a similar theme: the field of statistics was suffering from an image problem and needed to redefine itself in order to meet the challenges of a variety of existential threats. These threats included the rise of computational methods and large amounts of data, the emergence of non-traditional predictive methods and areas of research that were taking the limelight from statistics (i.e. Breiman's algorithmic models and data mining), and an unflattering public image. Interestingly, given the lack of reference to the Tokyo school, many of the proposed responses included expanding the scope of statistics to encompass these new methods and to rebrand the field as "data science." Frequently associated with this suggestion was a concern for updating university curricula for teaching statistics and, apparently for the first time, the creation of a new kind of statistician, the "data scientist"—recalling the mythical figure of the "whole statistician" discussed by Tukey in the 1970s. In these exhortations to the community of statisticians, the data scientist emerged as the "new man" of a reborn statistical science, one that who would overcome the field's crisis of recognition.

The first of these exhortations appears to have come in 1985 from the American statistician C. F. Jeff Wu in a lecture delivered at the Institute of Systems Science at the Chinese Academy of Sciences:

> I have an idea: Should China stop using the term "Statistics"? In Chinese, the word 'Statistics' literally means "general computation." However, this is not the statistics that you all have in mind. Should we call it Data Science (数据科学, Data Science)? This term more accurately reflects our work. Right now, people think that statistics is just the work done by statistical bureaus. Their work is mainly collecting data and performing preliminary organization. This aligns with the term "Statistics," but this is only a part of what statistics is about. If China wants to catch up and surpass others quickly, one shortcut would be to innovate in terminology first (laughter). In this way, when we introduce ourselves to others, we would not call ourselves statisticians (statisticians), but Data Scientists (Data Scientists) (C. F. J. Wu 1986 [1985]: 5; translated by ChatGPT).

More than a decade later, Wu echoed this idea in a lecture delivered at University of Michigan entitled "Statistics = Data Science?" (C. F. Jeff Wu 1997).

Regarding image, he pointed out that statisticians were perceived as either accountants or involved with simple descriptive statistics—and prone to lying with these statistics, as the saying goes—when in fact their work comprised everything from data collection to modeling and analysis to solving problems and making decisions. As a remedy, he implored his colleagues to "think big" and embrace the changes and challenges that we have seen already—the rise of large and complex data sets, the use of neural networks and data mining methods, and the emergence of new fields such as computer vision. In addition to suggesting a name change for the field, he appears to have been the first to suggest a name change of the role of statistician to data scientist, along with a college curriculum to that would embrace all phases of data science and be profoundly interdisciplinary.

This concern for image and a solution in an expanded role of statistics under the name "data science" appears elsewhere in the United States at this time. Jon R. Kettenring, in his Presidential Address to the American Statistical Association in 1997—three months before Wu gave his lecture—said this:

> Looking ahead, image reconstruction must be one of our top priorities. It must be understood that *statistics is the data science of the 21st century*—essential for the proper running of government, central to decision making in industry, and a core component of modern curricula at all levels of education. I would like to see ASA make image reconstruction a part of its strategic plan. And I suspect we may need some professional help if we are to succeed (Kettenring 1997: 1230; emphasis added).

In this talk, Kettenring argued that statistics, in professional practice and in education, needed to embrace topics in computer science, including "databases and database management, algorithm design, computational statistics, artificial intelligence and machine learning" (1230). Of interest here is the curious reversal of precedence between the two fields; whereas Wu asserted that statistics should expand its scope and name itself "data science," because it "is likely the remaining good name reserved for us" (Wu 1997: slide 12), Kettering's language implies the prior existence of data science as something that statistics should appropriate and encompass, as the rightful heir to its associated practices. Given the changes he suggested, in Kettering's usage of data science included much of computer science, which is consistent with the definition that developed in the 1960s and '70s.

Practical plans for revamped curricula to train this new kind of statistician—the data scientist—appear after these appeals. The most well-known is found in Cleveland's essay, "Data Science: An Action Plan for Expanding the Technical Areas of the Field of Statistics" (Cleveland 2001), even though he used the term "data analyst" as his target student. Consistent with previous definitions of data science as statistics augmented by computational methods and tools, he proposed a curriculum comprising six areas, along with percentages denoting the amount of time and resources that should be devoted to each: Multidisciplinary Investigations (25%), Models and Method for Data (20%), Computing with Data (15%), Pedagogy (15%), Tool Evaluation (5%), and Theory (20%). This distribution is more or less consistent with the definitions we have seen proposed by other statisticians, including the Tokyo school. As a measure of how radical this suggestion was, consider Donoho's remarks, made over a decade and a half later:

> Several academic statistics departments that I know well could, at the time of Cleveland's publication, fit 100% of their activity into the 20% Cleveland allowed for theory. Cleveland's article was republished in 2014. I cannot think of an academic department that devotes today 15% of its effort on pedagogy, or 15% on computing with data. I can think of several academic statistics departments that continue to fit essentially all their activity into the last category, theory (Donoho 2017: 750).

Radical as it was, from the perspective of Breiman's essay the curriculum was fairly conservative. The area of "models and methods," for example, focused exclusively on data models, as opposed to algorithmic ones, while the "computing with data" area focused narrowly on the infrastructure to support these models. Cleveland's bias toward traditional statistical modeling is made clear in his explanation of it:

> The data analyst faces two critical tasks that employ statistical models and methods: (1) specification—the building of a model for the data; (2) estimation and distribution—formal, mathematical-probabilistic inferences, conditional on the model, in which quantities of a model are estimated, and uncertainty is characterized by probability distributions (22).

That the one is subordinate to the other is evident in Cleveland's remark: "A collection of models and methods for data analysis will be used only if the collection is implemented in a computing environment that makes the models and methods sufficiently efficient to use" (23). He did make a passing reference to algorithmic models, but his example came from their use to support stochastic modeling:

> Historically, the field of data science has concerned itself only with one corner of this large domain [i.e. computing with data]—computational algorithms. Here, even though effort has been small compared with that for other areas, the impact has been large. One example is Bayesian methods, where breakthroughs in computational methods took a promising intellectual current and turned it into a highly practical, widely used general approach to statistical inference (23).

If it is not made clear from this passage that he did not have the wider field of data mining and knowledge discovery in mind, the following passage does:

> Computer scientists, waking up to the value of the information stored, processed, and transmitted by today's computing environments, have attempted to fill the void. Once current of work is data mining. *But the benefit to the data analyst has been limited, because knowledge among computer scientists about how to think of and approach the analysis of data is limited*, just as the knowledge of computing environments by statisticians is limited (23; emphasis added).

So, Cleveland continued the practice of the Tokyo school, *contra* Breiman, to dismiss the contributions of data miners, for their lack of training in traditional statistics, or lack of attention to data design, and to confine the role of computational expertise to knowledge of "environments."

Regarding the argument being made here, that the term data science was in continuous circulation since the 1960s, and not independently coined in various contexts, Cleveland's remarks, similar to those of Kettenring, indicate that the term was known to statisticians, even as they sought to appropriate it for their own purposes. The sentence—"Historically, the field of data science has concerned itself only with one corner of this large domain …"—strongly implies awareness of prior usage, as well as a definition that aligns with what we have called classical data science. In any case, as we have seen, CODATA, representing classical data science, began publishing the *Data Science Journal* in 2001.

The task of educating data scientists was also addressed at this time at a workshop sponsored by the American Statistical Association in 2000 on the topic of undergraduate education in statistics. In a report of the proceedings published in the *American Statistician*, the following understanding of data science was expressed:

> … what is needed is a broader conception of statistics, a conception that includes data management and computer skills that assist in managing, exploring, and describing data. The terms "data scientist" or "data specialist" were suggested as perhaps more accurate descriptions of what should be desired in an undergraduate statistics degree. Data specialists would be concerned with the "front end" of a data analysis project: designing and managing data collection, designing and managing databases, manipulating and transforming data, performing exploratory and "basic" analysis (Higgins 1999). Data scientists (or specialists) a might share some course work with computer science majors, but where a computer scientist studies compilers and assembly language, a data scientist studies data analysis and statistics (Bryce et al. 2001: 12; citation in original).

The reference to Higgins is from a paper where he asserted the need for statisticians to pay more attention to "the non-mathematical part of statistics," so that undergraduate programs may respond to the "explosion in the amount of data available to society" (Higgins 1999: 1). In this area he included "designing scientific studies in a team-oriented environment, ensuring protocol compliance, ensuring data quality, managing the storage/transmission/retrieval of data, and providing descriptive and graphical analyses of data" (1). Here, again, we see a definition of data science as an improved version of statistics, in response to the persistent condition of data impedance, that would include data management and computing skills (and data design)—but not the methods of data mining. Consistent with the implied structural relationship between statistics and data science, data science was sometimes described as a part of statistics. Tellingly, both Bryce, et al., and Higgins equivocated on the use of data scientist, and suggested "data specialist" as an alternate name, perhaps so as not to overshadow the role of the data analyst. In any case, the choice of term was motivated by marketing; as Higgins wrote:

> Guttman expressed the opinion that the term statistics carries such a negative connotation that it might be wise to rename our departments something like "Department of Data Science" or "Department of Information and Data Science." In this vein, I have suggested the term "data specialist" (Higgins 1999).[14]

## 3.5. 2005: Data Science in the Sciences

As some members of the statistics community presented plans to incorporate data science into their field, the terms "data science" and "data scientist" nevertheless continued to be used in the classical sense of the science of data in the service of science. Indeed, by 2005, the role of data scientist had become sufficiently developed within the scientific community that it appeared as a central element in a report from the US National Science Foundation (NSF), "Long-Lived Digital Data Collections: Enabling Research and Education in the 21st Century" (Simberloff et al. 2005). The report defines the role in specific terms:

DATA SCIENTISTS

The interests of data scientists—the information and computer scientists, database and software engineers and programmers, disciplinary experts, curators and expert annotators, librarians, archivists, and others, who are crucial to the successful management of a digital data collection—lie in having their creativity and intellectual contributions fully recognized. In pursuing these interests, they have the responsibility to:

- conduct creative inquiry and analysis;

- enhance through consultation, collaboration, and coordination the ability of others to conduct research and education using digital data collections;

- be at the forefront in developing innovative concepts in database technology and information sciences, *including methods for data visualization and information discovery*, and applying these in the fields of science and education relevant to the collection;

- implement best practices and technology;

- serve as a mentor to beginning or transitioning investigators, students and others interested in pursuing data science; and design and implement education and outreach programs that make the benefits of data collections and digital information science available to the broadest possible range of researchers, educators, students, and the general public.

Almost all long-lived digital data collections contain data that are materially different: text, electro-optical images, x-ray images, spatial coordinates, topographical maps, acoustic returns, and hyper-spectral images. In some cases, it has been the data scientist who has determined how to register one category of representation against another and how to cross-check and combine the metadata to ensure accurate feature registration. Likewise, there have been cases of data scientists developing a model that permits representation of behavior at very different levels to be integrated. *Research insights can arise from the deep understanding of the data scientist of the fundamental nature of the representation. Such insights complement the insights of the domain expert. As a result, data scientists sometimes are primary contributors to research progress.* Their contribution should be documented and recognized. One means for recognition is through publication, i.e., refereed papers in which they are among the leading authors (Simberloff et al. 2005: 26; emphases added).

This account of the role of data scientist demonstrates both the currency of the term and its adherence to the classical definition. Again, this usage stood in contrast to that developed in the statistics community for its emphasis on the creative role of data curation and representation and its sympathetic view toward knowledge discovery. It is also worth noting the normative intent of the definition—the report described the role of data scientist as both heterogenous—comprising a wide array of knowledge workers from computer scientists to librarians—and undervalued. The report sought to correct this condition. As evidence for its influence, consider that Purdue University's Distributed Data Curation Center (D2C2), founded in 2007 as "a research center that would connect domain scientists, librarians, archivists, computer scientists, and information technologists" to address "the need by researchers for help in discovering, managing, sharing, and archiving their research data," included "a full-time Data Research Scientist, a position based on the data scientist role" as described in the report (Witt 2008: 199).

---

[14] Higgins here referred to a quote from Guttman found in a paper by Kettring (Kettring 1997a).

The NSF report drew on the established usage of the term in the scientific community. During this period there appeared numerous instances of the term "data scientist" in popular media that are consistent with the classical definition of data science. For example, in 2008 *The Times* of London published a piece that quoted "Nathan Cunningham, 36, data scientist, British Antarctic Survey":

> When I am on the ship I am part of a team of scientists collecting data about everything from the biomass in the ocean to the weather patterns. … Our monitoring equipment is always on and sends us 180 pieces of information every second. *My role is to make sure that each person can find the exact data that they want among all this, so I write programs to help them to do this*. Another one of my field responsibilities is getting the information that we collect back to Cambridge via satellite link so that other researchers can use the data (Chynoweth 2008; emphasis added).

Other stories about data scientists were reported in news media touting the work of local universities, such as at Brigham Young University and Rensselaer Polytechnic Institute (Harmon 2007; Targeted News Service 2008). The *New Scientist* posted job ads for data scientists as far back as 1992 ("New Scientist" 1992; "New Scientist" 1995; "New Scientist" 1996; "New Scientist" 1999; "New Scientist" 2001). In some cases, the term was prefixed, as in "Clinical Data Scientist," "Marine Data Scientist," and "Senior Data Scientist," but in others it was not.

Alongside but contrary to plans for data science curricula from the statisticians' perspective, computer scientists and scientists in disciplines that had long been engaged with data impedance, such as physics and astronomy, began to outline requirements for data science to become a mature field. In "Data Science as an Academic Discipline," published in CODATA's *Data Science Journal*, Irish computer scientist F. Jack Smith (OBE) argued for data science to develop its own peer-reviewed body of knowledge, in the form of refereed journals and textbooks, on the premise that "[o]nce a body of literature is in place, academic courses can begin at universities" (Smith 2006: 164). Consistent with the journal's source, Smith's definition of data science was different to that proposed by Wu and Cleveland. The following historical perspective makes this clear:

> To be taken seriously, any discipline needs to have endured over time. Unlike computers, scientific data has a long history. Without astronomic data, Newton would not have discovered gravitation. Without data on materials, the Titanic would not have been built, and with good data on the location of icebergs, it might not have sunk! Data then consisted of tables of facts and quantities found in textbooks and journals, but data science did not yet exist. *Then computers and mass storage devices became available, and the first databases were designed holding scientific data. Data science was born soon afterwards, about 1966, when a few far seeing pioneers formed CODATA.*

> Data science has developed since to include the study of the capture of data, their analysis, metadata, fast retrieval, archiving, exchange, *mining to find unexpected knowledge and data relationships*, visualization in two and three dimensions including movement, and management. Also included are intellectual property rights and other legal issues.

> Data science, however, has become more than this, something that the pioneers who started CODATA could not have foreseen; data has ceased being exclusively held in large databases on centrally located main frames but has become scattered across an internet, instantly accessible by personal computers that can themselves store gigabytes of data. *Therefore, the nature and scope of much scientific and engineering data and, in consequence, of much scientific research has changed.* Measurement technologies have also improved in quality and quantity with measurement times reduced by orders of magnitude. Virtually every area of science, astronomy, chemistry, geoscience, physics, biology, and engineering is also becoming based on *models dependent on large bodies of data, often terabytes, held in large scientific data systems* (163; emphases added).

This view, close to that espoused here, locates data science in the historically specific emergence of networked, computational databases—what has been called the datasphere (Garfinkel 2000; Alvarado and Humphreys 2017). This emphasis on the dependence of models on this infrastructure represents what seems to be a distinct view to that of the statistician, who tends to regard these developments as exogenous to her engagement with data. Put another way, for the statistician, the historical shift from data to databases—from print to digital modes of communication—is often represented as a difference in degree, but for the scientist, who produces and lives among these data, it is a difference in kind. This difference in perspective was not without epistemic consequences. For one, Smith's definition clearly included data mining. For another, just as data mining had been excluded from the statistician's definition of data science, so too was a concern for databases excluded from what was considered worthwhile scientific work:

> I recall being a proud young academic about 1970; I had just received a research grant to build and study a scientific database, and I had joined CODATA. I was looking forward to the future in this new exciting discipline when the head of my department, an internationally known professor, advised me that data was "a low level activity" not suitable for an academic. I recall my dismay. What can we do to ensure that this does not happen again and that data science is universally recognized as a worthwhile academic activity? Incidentally, I did not take that advice, or I would not be writing this essay, but moved into computer science. I will use my experience to draw comparisons between the problems computer science had to become academically recognized and those faced by data science (Smith 2006: 163).[15]

Further evidence for the divergent conceptions of data science held by statisticians and scientists during this period appears in an ambitious position paper prepared for the 2010 Astronomy and Astrophysics Decadal Survey, written to "address the impact of the emerging discipline of data science on astronomy education" (Borne et al. 2009). Building on Smith's conception of both science and data science, the report cited the usual concerns with data impedance—the gap between information and data on the one hand and knowledge and understanding on the other, produced by "information explosion" and "exponential data deluge." As a response, the authors proposed to redefine science as fundamentally data-driven and dependent upon computational technologies. Indeed, in their four-part model, data were depicted as central, as the fourth node within a triangle consisting of "Sensor," "HPC," and "Model." The result was a conception of science in which data science would participate as a first-class member:

> The emerging confluence of new technologies and approaches to science has produced a new Data-Sensor-Computing-Model synergism. This has been driven by numerous developments, including the information explosion, the development of dynamic intelligent sensor networks …, the acceleration in high performance computing (HPC) power, and advances in algorithms, models, and theories. *Among these, the most extreme is the growth in new data.* The acquisition of data in all scientific disciplines is rapidly accelerating and causing a nearly insurmountable data avalanche [3 (Bell, Gray, and Szalay 2007)]. Computing power doubles every 18 months (Moore's Law), corresponding to a factor of 100 in ten years. The I/O bandwidth (into and out of our systems, including data systems) increases by 10% each year—a factor 3 in ten years. By comparison, data volumes appear to double every year (a factor of 1,000 in ten years). Consequently, as growth in data volume accelerates, especially in the natural sciences (where funding certainly does not grow commensurate with data volumes), we will fall further and further behind in our ability to access, analyze, assimilate, and assemble knowledge from our data collections—unless we develop and apply increasingly more powerful algorithms, methodologies, and approaches. *This requires a new generation of scientists and technologists trained in the discipline of data science* [4 (Shapiro et al. 2006)] (1–2; emphases and citations added).

---

[15] Based on the affiliations cited in his three publications between 1960 and 1969, the setting for Smith's story was the School of Physics and Applied Mathematics at the Queen's University of Belfast in Northern Ireland.

The inclusion of data mining in this conception is clear:

> We see the data flood in all sciences (e.g., numerical simulations, high-energy physics, bioinformatics, geosciences, climate monitoring and modeling) and outside of the sciences (e.g., banking, healthcare, homeland security, drug discovery, medical research, retail marketing, e-mail). *The application of data mining, knowledge discovery, and e-discovery tools to these growing data repositories is essential* to the success of our social, financial, medical, government, and scientific enterprises. (2; emphasis added)

Although Cleveland's action plan was cited in this report, as evidence that "data science is becoming a recognized academic discipline" (3), it is clear his definition of data science was not adopted. Instead, a conception of data science that included data mining and which would play a central role in the scientific enterprise was more reflective of the view expressed in Microsoft's contemporary and influential publication, *The Fourth Paradigm: Data-Intensive Scientific Discovery* (Hey, Tansley, and Tolle 2009a). Although the various authors did not use the term "data science" at all, the role played by data, data technologies, and specifically data mining, were highlighted throughout. To anticipate what follows, the fourth paradigm concept would later become one of the dominant, competing definitions of data science once the term is popularized after 2010.

If Microsoft's report did not use the term, other organizations cited within the report did. For example, what was then known as the Joint Information Systems Committee (JISC), established by the UK in 1993 to provide guidance to networking and information services to the entire kingdom's higher education sector, sponsored a report "to examine and make recommendations on the role and career development of data scientists and the associated supply of specialist data curation skills to the research community" (Swan and Brown 2008: 1). Aware of the semantic confusion surrounding the term by this time, the report offered this helpful clarification of roles:

> The nomenclature that currently prevails is inexact and can lead to misunderstanding about the different data-related roles that exist. … We distinguish four roles: data creator, data scientist, data manager and data librarian. We define them in brief as follows:
>
> - Data creator: researchers with domain expertise who produce data. These people may have a high level of expertise in handling, manipulating and using data
>
> - Data scientist: people who work where the research is carried out—or, in the case of data centre personnel, in close collaboration with the creators of the data—and may be involved in creative enquiry and analysis, enabling others to work with digital data, and developments in data base technology
>
> - Data manager: computer scientists, information technologists or information scientists and who take responsibility for computing facilities, storage, continuing access and preservation of data
>
> - Data librarian: people originating from the library community, trained and specialising in the curation, preservation and archiving of data

In practice, there is not yet an exact use of such terms in the data community, and the demarcation between roles may be blurred. It will take time for a clear terminology to become general currency.

Data science is now a topic of attention internationally. In the USA, Canada, Australia, the UK and Europe, developments are occurring. It is notable that the vision in all these places is that data science should be organised and developed on a national pattern rather than relying on piecemeal approaches to the issues (1).

These definitions, which expand our scope to include the wider division of labor within which data work took place, are illuminating. They show that even as late as 2008, at precisely the time when a new usage of data scientist would emerge from Silicon Valley, the role was still more closely associated with the classical definition than with the newer definitions proposed by the Tokyo school, Wu, and Cleveland. Again, the salient difference concerns the role of the data scientist (or specialist) relative to the liminal site of data creation at the heart of empirical science: the distinguishing features of the definition given above are that the data scientist works in "close collaboration with the creators of the data," "where the research is carried out," and "may be involved in creative enquiry and analysis." Indeed, the two roles, of researcher and data scientist, may be combined in one person. We may be sure that "creative enquiry" here refers to more than the kind of data modeling performed by Breiman's data modeling culture. To make the separation between statistician and data scientist clearer, consider the following remark, made in reference to one perspective on whether data science should be taught at the undergraduate level: "data skills should be viewed as a fundamental part of the education of undergraduates in the same way as basic statistics, laboratory practices and methods of recording findings are" (24).

It is worth noting the curious appearance of the term "dataology" as this time, spelled differently than Naur's "datology," in the work of Zhu, et al. as a synonym for data science. Apparently unaware of Naur, these authors proposed a new science of data in response to data impedance ("data explosion") that would focus on what they termed "data nature":

The essence of computer applications is to store things in the real world into computer systems in the form of data, i.e., *it is a process of producing data*. Some data are the records related to culture and society, and others are the descriptions of phenomena of universe and life. The large scale of data is rapidly generated and stored in computer systems, which is called *data explosion*. *Data explosion* forms *data nature* in computer systems. To explore data nature, new theories and methods are required. In this paper, we present the concept of data nature and introduce the problems arising from data nature, and then we define a new discipline named *dataology* (also called *data science* or *science of data*), which is an umbrella of theories, methods and technologies for studying data nature. The research issues and framework of *dataology* are proposed (Zhu, Zhong, and Xiong 2009: abstract; emphases in original).

This definition was consistent with the classical definition and indeed echoed the concerns of the US Department of Defense to define data decades earlier, as a prerequisite to developing technologies to process and manage it. It is also worth noting the change in understanding of data impedance at this juncture; whereas originally the focus was on the overproduction of data by sources ranging from scholarly communication to satellite signals, in relation to machinery to process it, by this time it referred to vast amounts of data collected in databases—the machinery developed to manage data. This parallels the shift in focus we saw in the Tokyo school, from raw data to data in databases. As the locus of impedance changed, so too did the focus of data science (in this usage). For Zhu, et al., data nature referred explicitly to data in databases and computer systems, and their concern was to understand the relationship between data nature and real nature. Again, this shift is consistent with the classical definition as well as Naur's; the focus is on the epistemological dimensions of data, data as a form of representation, as found in computational machinery. In contrast to Nau, however, the relationship between data and the world is considered central. From this perspective, data mining is regarded as a kind of data science:

> The appearance of data mining technology … means that people began to study the laws aiming at data in computer systems. In the field of Internet, more and more researches focus on network behavior, network community, network search, and network culture. Because of the accumulation of data, newly disciplines, such as bioinformatics and brain informatics, are also typical dataology centric research areas. For instance, DNA data in bioinformatics are the data that describe natural structures of life, based on which we can study life using computers (153).

In other words, not only is data mining consistent with data science in this view, it is central to it. More recently, the authors situate this definition within the array of definitions that currently characterize the ambiguous nature of the field and which motivate the current essay (Zhu and Xiong 2015). Consistent with their focus on data as it exists in databases, the authors focus on the role of the Internet and social media in constituting the field of data science. This is a perspective we will revisit.

During the first two decades following Berners-Lee's invention of the Web, the term data science emerged from the long tail of usage, where it had resided essentially since Naur, to become the sign of new kind of statistics. This new statistics would overcome the limitations of a field that had lost its way amidst the rise of computational data processing technologies and of what would eventually be called "big data," a term that we may take to be a synonym for the condition of data impedance associated with the use of the term data science since the 1960s. Ironically, big data was the product of what we might call the first wave response to data impedance; database technologies were developed to contain and manage the oft mentioned "data deluge," and these in turned produced another flood, of software and enormous caches of aggregated data. In addition, they produced a nemesis to statistics—the field of data mining.

Yet throughout out this period, classical data science persists and paradoxically becomes stronger, perhaps in response to the use of the term among statisticians. We see that even up to the eve of the next milestone, data science was widely understood to refer to work associated with data processing, the theory and practice associated with data, especially in the context of scientific research data. This was the understanding of data science implied by CODATA's journal. We also note that the classical definition, when it was articulated, was consistently inclusive of the work of data miners. For their part, these new workers did not appear to need the term; the phrase "knowledge discovery (in databases)," referring to the framing context of activity, was sufficient to capture their understanding of their work (Fayyad et al. 1996).

Notably, during this period the term "data scientist" emerged as well, in both statistical and classical contexts. Its appearance reflected a concern for data science as an emerging profession, complete with educational requirements. In the statistician's usage, this position was ambivalently placed within the general division of labor, as either a synthetic "new man" figure that would encompass data analysis, or else as a "data specialist," an adjunct to the more primary work of the data analyst. In any case, the appearance of this grammatical variant indicates a transformation in the social context of usage: data science had moved from being an abstract concern to a widely distributed and embedded activity.

### 3.6. 2008: The Sexy Science

Around 2008, a decisive shift in the meaning of "data science" (and "data scientist") took place. After nearly a half century of development, in which two broad and consistent usages had emerged—the classical and the statistical—the term was applied in a context that launched it into the public eye and increased its circulation by orders of magnitude.[16] This context was the social media corporation of the Web 2.0 era, itself the inheritor of two crowning achievements of data processing engineering—the database and the Internet—catalyzed by Berners-Lee's invention of a global hypertext system. According to what is perhaps the most popular article on the topic, *Harvard Business Review*'s "Data Scientist: The Sexiest Job of the 21st Century," the term "data scientist" was "coined in 2008 by ... D.J. Patil, and Jeff Hammerbacher, then the respective leads of data and analytics efforts at LinkedIn and Facebook" (Davenport and Patil 2012).[17] Although this claim is obviously false, it is apparently correct in having identified the first usages of the term in this new context.

Consistent with this claim, in 2009 Hammerbacher published an essay recounting his experience as a data scientist at Facebook entitled "Information Platforms and the Rise of the Data Scientist" (Hammerbacher 2009), where he explained the motivation for adopting the term:

> At Facebook, we felt that traditional titles such as Business Analyst, Statistician, Engineer, and Research Scientist didn't quite capture what we were after for our team. The workload for the role was diverse: on any given day, a team member could author a multistage processing pipeline in Python, design a hypothesis test, perform a regression analysis over data samples with R, design and implement an algorithm for some data-intensive product or service in Hadoop, or communicate the results of our analyses to other members of the organization in a clear and concise fashion. To capture the skill set required to perform this multitude of tasks, we created the role of "Data Scientist" (84).

Although the work in this description appears evenly divided among engineering and statistical tasks, Hammerbacher's narrative actually focused entirely on efforts to manage the data impedance problem that social media company (and others like it) faced at the time he was hired in 2006, just after they opened their doors to non-college students. It is a story of how a small group within the company moved away from a MySQL data warehouse, which literally ceased to function under the volume of the company's data, and eventually to a new platform based on Hadoop and Hive, in order to perform the standard tasks of extracting, transforming, and loading data (ETL) and building an information retrieval platform for analysts in the company. In addition to their massive scale, these data were also textual and social in nature, putting them in the same category as the complex data types the Air Force faced years ago.

Significantly, Hammerbacher emphasized Facebook's adoption of Google's machine learning based approach, captured in the phrase "the unreasonable effectiveness of data" and explained in an essay with that title (Halevy, Norvig, and Pereira 2009). In this approach, the size of data is considered more important than the sophistication of models—"invariably, simple models and a lot of data trump more elaborate models based on less data" (9; also quoted by Hammerbacher).[18] In essence, then, the new role of data scientist at Facebook was close to that of the data scientist for the British Antarctic Survey, i.e. the classical, not the statistical definition, with the added focus on machine learning. Indeed, Hammerbacher appears to have recognized the provenance of the term:

> Outside of industry, I've found that grad students in many scientific domains are playing the role of the Data Scientist. One of our hires for the Facebook Data team came from a bioinformatics lab where he was building data pipelines and performing offline data analysis of a similar kind. The well-known Large Hadron Collider at CERN generates reams of data that are collected and pored over by graduate students looking for breakthroughs (Hammerbacher 2009: 84).

---

[16] According to the Google Books NGram Viewer, using the English (2019) corpus with a smoothing factor of 3, the corpus frequency of the bigram "data scientist" (case insensitive) goes from 0.0000000613% in 2008 to 0.0000035018% in 2012 and 0.0000118327% in 2016. These are increases from 2008 of around 57 and 193 for 2012 and 2016 respectively. By comparison, frequency actually *decreases* by .6 from 2000 (.0000000876%) to 2008, although the difference here is probably not significant statistically; the frequency is essentially flat. The trend is similar for "data science," with increases of 11 and 56 times from 2008 to 2012 and 2016 respectively.

[17] The title of this article was adapted from a phrase used in 2008 by Hal Varian, chief economist at Google. In an interview with McKinsey's James Manyika, Varian quipped: "I keep saying the sexy job in the next ten years will be statisticians" (McKinsey & Company 2009).

It seems likely that the phrase data scientist, as it was understood then by the scientific community, was transferred to this new domain by Hammerbacher's contact with some of its members.

At around the same time that Hammerbacher and Patel are said to have coined the phrase data scientist, in October 2008 Hal Varian, chief economist at Google, gave an interview to McKinsey's James Manyika in which he uttered the famous phrase, "I keep saying the sexy job in the next ten years will be statisticians [sic]" (McKinsey & Company 2009). The interview was published on new year's day of 2009 and was immediately processed by the blogosphere. Nathan Yau, who has a Ph.D. in statistics from UCLA and runs the blog *FlowingData*, devoted to data visualization, was quick to qualify Varian's use of the term "statisticians":

> … if you went on to read the rest of Varian's interview, you'd know that by *statisticians*, he actually meant it as a general title for someone who is able to extract information from large datasets and then present something of use to non-data experts (Yau 2009a).

Here's what Varian actually said:

> I think statisticians are part of it, but it's just a part. *You also want to be able to visualize the data, communicate the data, and utilize it effectively.* But I do think those skills—of being able to access, understand, and communicate the insights you get from data analysis—are going to be extremely important. Managers need to be able to access and understand the data themselves (McKinsey & Company 2009; emphases added).

In a follow-up post, "Rise of the Data Scientist," Yau expands on his revision to Varian's remarks by incorporating the comments of fellow blogger, Michael Driscoll of *Dataspora*, who also responded to the McKinsey piece in a post entitled "The Three Sexy Skills of Data Geeks." Echoing Yau, Driscoll wrote:

> I believe that the folks to whom Hal Varian is referring are not statisticians in the narrow sense, but rather people who possess skills in three key, yet independent areas: statistics, data munging, and data visualization (Driscoll 2009).

Yau connects Driscoll's insight to Ben Fry's fertile concept of "computational information design" (Fry 2004), which maps the fields of computer science, mathematics, statistics, data mining, graphic design, and human-computer interaction onto a data processing pipeline— acquire, parse, filter, mine, represent, refine, and interact. Whereas Driscoll called the role that integrates these fields "statisticians or data geeks," tellingly equivocating, Yau used the term data scientist:

> And after two years of highlighting visualization on FlowingData, it seems collaborations between the fields are growing more common, but more importantly, computational information design edges closer to reality. We're seeing *data scientists* – people who can do it all – emerge from the rest of the pack.
>
> . . .
>
> They have a combination of skills that not just makes independent work easier and quicker; it makes collaboration more exciting and opens up possibilities in what can be done. Oftentimes, visualization projects are disjoint processes and involve a lot of waiting. Maybe a statistician is waiting for data from a computer scientist; or a graphic designer is waiting for results from an analyst; or an HCI specialist is waiting for layouts from a graphic designer.
>
> Let's say you have several data scientists working together though. There's going to be less waiting and the communication gaps between the fields are tightened.
>
> How often have we seen a visualization tool that held an excellent concept and looked great on paper but lacked the touch of HCI, which made it hard to use and in turn no one gave it a chance? How many important (and interesting) analyses have we missed because certain ideas could not be communicated clearly? *The data scientist can solve your troubles* (Yau 2009b; emphasis added).

---

[18] By "simple," the authors meant few assumptions about the nature of the data, such as are required by data models that posit parametric distributions or causal relations among features. The usage is ironic, since one of the differences between the inferential models preferred by traditional statisticians and predictive models is that the former are chosen for their parsimony and interpretability, whereas the latter notoriously have too many terms to interpret.

Yau's definition of data scientist is consistent with that given in the *HBR* article written four years later. There the authors described data scientists as those who "make discoveries while swimming in [the deluge of] data," "bring structure to large quantities of formless data and make analysis possible," "identify rich data sources, join them with other, potentially incomplete data sources, and cleaning the resulting set," "help decision makers shift from ad hoc analysis to an ongoing conversation with data," "are creative in displaying information visually and making the patterns they find clear and compelling," "advise executives and product managers on the implications of the data for products, processes, and decisions,' and so on. Replace the roles of decision maker, executive, and product manager with scientist and engineer, and the definition is remarkably consistent with the classical definition as it had been developed in the years leading up to this shift.

Yau and Driscoll's response to Varian are notable because they demonstrate how new the terms data science and data scientist were to the general public at the time, and the manner in which these terms transitioned from a narrow community of discourse to a much larger one. Varian and Driscoll used the term statistician, but both had to qualify them significantly. That "data scientist" was not yet mainstream in 2009 can be seen in a *New York Times* article also written in response to the McKinsey interview, entitled "For Today's Graduate, Just One Word: Statistics" (Lohr 2009). Here, the author was compelled to use the expression "Internet-age statistician" to name the role described by Varian, and to qualify this usage in a manner similar to Varian:

> Though at the fore, statisticians are only a small part of an army of experts using modern statistical techniques for data analysis. Computing and numerical skills, experts say, matter far more than degrees. So the new data sleuths come from backgrounds like economics, computer science and mathematics.

We might note here an important difference between how the two bloggers, closer to the reality being described, and the established-media journalist represented the new development heralded by Varian: whereas Lohr saw an expanded division of labor, Yau and Driscoll envisioned an entirely new role, something akin to the "whole statistician" described above, although combining different elements. In any case, we see that it is at this precise point that the term "data scientist" begins to be used in the new context later described by the *HBR*.

In September 2010, two short but highly influential blog posts appeared that sought to codify this conception of data science, which had by then reached the status of buzz word among participants of the technology conference circuit. The first was Drew Conway's "The Data Science Venn Diagram," which defined the field as the intersection of three areas: "hacking skills, math and stats knowledge, and substantive expertise" (Conway 2010). Conway's post followed his attending an "unconference to help O'Reily [sic] organize their upcoming Strata conference," where he detected "the utter lack of agreement on what a curriculum on this subject would look like." His rationale for the three areas was the following:

> … we spent a lot of time talking about "where" a course on data science might exist at a university. The conversation was largely rhetorical, as everyone was well aware of the inherent interdisciplinary nature of the [sic] these skills; but then, why have I highlighted these three? First, none is discipline specific, but more importantly, each of these skills are on their own very valuable, but when combined with only one other are at best simply not data science, or at worst downright dangerous.

Of interest here is, first, the need to define a curriculum for what was perceived to be a new field, which echoed previous efforts and presaged the academic response that would eventually follow, and second, that Conway's was essentially the classical definition applied to the context of what we might call information capitalism, the target audience of O'Reilly's conference. Although the role of statistics is emphasized, Conway reduces its importance to "knowing what an ordinary least squares regression is" and goes on to assert that "data plus math and statistics only gets you machine learning." In other words, Conway's definition is closer to Breiman's culture of algorithmic modeling than it is to that of data modeling. This is corroborated by the fact that by *data* Conway meant "a commodity traded electronically," i.e. that which is found in databases and shared over networks, as opposed to that which is collected intentionally by designed experiments (A/B testing notwithstanding).

The second post was Mason and Wiggins' "A Taxonomy of Data Science," which was motivated by the need to make sense of the newly circulated term:

> Both within the academy and within tech startups, we've been hearing some similar questions lately: Where can I find a good data scientist? What do I need to learn to become a data scientist? Or more succinctly: What *is* data science? (Mason and Wiggins 2010)

In contrast to Conway's structural model, Mason and Wiggins propose a processual one, based on "what a data scientist does, in roughly chronological order: Obtain, Scrub, Explore, Model, and iNterpret (or, if you like, OSEMN, which rhymes with possum)." In this model, most of the activities normally associated with the classical definition of data science—as listed in the *HBR* piece, for example—find a place. A distinguishing feature of this definition is the modeling phase, which they characterized as follows:

> Whether in the natural sciences, in engineering, or in data-rich startups, often the 'best' model is the most predictive model. E.g., is it 'better' to fit one's data to a straight line or a fifth-order polynomial? Should one combine a weighted sum of 10 rules or 10,000? One way of framing such questions of model selection is to remember why we build models in the first place: to predict and to interpret. While the latter is difficult to quantify, the former can be framed not only quantitatively but empirically. That is, armed with a corpus of data, one can leave out a fraction of the data (the "validation" data or "test set"), learn/optimize a model using the remaining data (the "learning" data or "training set") by minimizing a chosen loss function (e.g., squared loss, hinge loss, or exponential loss), and evaluate this or another loss function on the validation data. Comparing the value of this loss function for models of differing complexity yields the model complexity which minimizes generalization error. The above process is sometimes called "empirical estimation of generalization error" but typically goes by its nickname: "cross validation." Validation does not necessarily mean the model is "right." As Box warned us, "all models are wrong, but some are useful". Here, we are choosing from among a set of allowed models (the `hypothesis space`, e.g., the set of 3rd, 4th, and 5th order polynomials) which model complexity maximizes predictive power and is thus the least bad among our choices.

Clearly, the authors sided with algorithmic modeling here; they argued for prediction over interpretation by citing methods that have become commonplace in the field. We also find here, for the first time, a clearly articulated pipeline of activity, echoing the partial sequences that appear in previous definitions. Again, it is worth noting what data meant in this context: data are to be obtained from preexisting sources, sometimes by scraping, and not produced. The skills required are far from those of the design-oriented data scientist of the Tokyo school:

> Part of the skillset of a data scientist is knowing how to obtain a sufficient corpus of usable data, possibly from multiple sources, and possibly from sites which require specific query syntax. At a minimum, a data scientist should know how to do this from the command line. e.g., in a UN*X environment. Shell scripting does suffice for many tasks, but we recommend learning a programming or scripting language which can support automating the retrieval of data and add the ability to make calls asynchronously and manage the resulting data. Python is a current favorite at time of writing (Fall 2010).

Note that the idea of a data analyst *looking* for "usable data" as a first resort is anathema to that approach, at least in principle.

In 2011, O'Reilly, whose role in the promotion of data science is worth its own investigation, produced a series of influential blog posts and reports that sought to codify and amplify the definition developed by Hammerbacher, Yau, Conway, Mason, and Wiggins.[19] The definition produced was consistent with the classical version, but strongly inflected by the new business context. For example, Loukides' in "What is Data Science?" described the field in terms consistent with what we have seen, focusing on scale, new database technologies, and machine learning in the pattern set by Google. However, in this discourse these elements are combined in the new concept of the "data product," a good or service that integrates surplus data to provide value to users:

> The web is full of "data-driven apps." Almost any e-commerce application is a data-driven application. There's a database behind a web front end, and middleware that talks to a number of other databases and data services (credit card processing companies, banks, and so on). But merely using data isn't really what we mean by "data science." A data application acquires its value from the data itself, and creates more data as a result. *It's not just an application with data; it's a data product.* Data science enables the creation of data products (Loukides 2011; emphasis added).

This emphasis on data products was echoed in Patil's essay, "Building data science teams," where the focus on data *applications* became essential to his definition of data scientist. Here he addressed the question, "What makes a data scientist?":

> When Jeff Hammerbacher and I talked about our data science teams, we realized that as our organizations grew, we both had to figure out what to call the people on our teams. "Business analyst" seemed too limiting. *"Data analyst" was a contender, but we felt that title might limit what people could do. After*

---

[19] The influence of O'Reilly on this history is worth its own essay.

*all, many of the people on our teams had deep engineering expertise.* "Research scientist" was a reasonable job title used by companies like Sun, HP, Xerox, Yahoo, and IBM. However, *we felt that most research scientists worked on projects that were futuristic and abstract, and the work was done in labs that were isolated from the product development teams.* It might take years for lab research to affect key products, if it ever did. Instead, *the focus of our teams was to work on data applications that would have an immediate and massive impact on the business.* The term that seemed to fit best was data scientist: those who use both data and science to create something new (Patil 2011; empases added).

The focus on data products at this point in history may be understood in light of Zuboff's thesis that Google invented the business model of "surveillance capitalism" around 2003, based on the extraction of "behavioral surplus," which was then exported to Facebook by Cheryl Sandberg in 2008 and became widespread after that (Zuboff 2019). Zuboff makes sense of the fact that in the O'Reilly papers Google's services were frequently presented as exemplary data products, as well as the fact that the role of data scientist emerged at Facebook during the year of Sandberg's arrival there. It also sheds light on the nature of the "immediate and massive impact" of data products described by Patel: the prototypical data product is Google's advertising auctioning platform, which, as a result of applying its massive amounts of behavioral data to predict user behavior, "produced a stunning 3,590 percent increase in revenue in less than four years" (Zuboff 2019: Ch. 3, Part VI). More generally, Zuboff sheds light on the practical context within which this new iteration of data science emerged: in the heart of the system of computer-mediated economic transactions described by Varian (Varian 2010). In the previous period, when data science was imagined to be located at the center of the infrastructure of data-driven science (as in the NSF report cited above), this setting is transferred to domain of global, Internet-mediated commerce. Thus, just as the phrase "data scientist" leapt from one context to another at this time, so did the infrastructural framework within which it made sense. Again, the meaning of data science remains relatively unchanged from the classical definition; what changes is the context.

By 2012, the terms data science and data scientist had trended widely in the media, in part due to the amplifying effects of the *HBR* article, which employed Varian's catchy quip but adopted Yau's characterization of its referent. Articles in sources such as the New York Times and Forbes regularly published stories on the demand for data scientists, profiles of data scientists and data-driven companies, and opinion pieces on its merits. In 2012, Forbes published a series of eight articles on "What is a Data Scientist," which featured interviews with self-identified data scientists from IBM, Tableau, LinkedIn, Amazon, and other places, demonstrating the currency of the term in industry at that time. In addition to media coverage, consulting firms such as Booz Allen and Gartner produced documents targeting at C-suite executives providing an overview of the field, including the definition of a data scientist (Herman et al. 2013; Laney 2012). Throughout these writings, the definitions provided were consistent with the newer meaning, roughly the combination of computer competency, data mining, statistical knowledge, communication and visualization skills, and business acumen. In addition, the terms big data and data science were highly correlated; sentences like "[d]ata scientists are the magicians of the Big Data era" were frequent (Miller 2013).

An important feature of the definition of data science in this period is its co-occurrence and close semantic association with the often-capitalized term "big data." The term was used to refer to both large amounts of data—retroactively identified with Laney's concept of "3D data," data with high "volume, velocity, and variety" (Laney 2001)—and to the assemblage of technologies and methods associated with these data. The following definition from ZDNet is typical:

"Big Data" is a catch phrase that has been bubbling up from the high performance computing niche of the IT market. Increasingly suppliers of processing virtualization and storage virtualization software have begun to flog "Big Data" in their presentations. What, exactly, does this phrase mean?

. . .

In simplest terms, the phrase refers to the tools, processes and procedures allowing an organization to create, manipulate, and manage very large data sets and storage facilities. Does this mean terabytes, petabytes or even larger collections of data? The answer offered by these suppliers is "yes" (Kusnetzky 2010).

Indeed, the terms big data and data science were also used interchangeably. This usage from a New York Times piece is representative:

*The field known as "big data"* offers a contemporary case study. The catchphrase stands for the modern abundance of digital data from many sources — the web, sensors, smartphones and corporate databases — that can be mined with clever software for discoveries and insights. Its promise is smarter, data-driven decision-making in every field. That is why data scientist is the economy's hot new job (Lohr 2014; emphasis added).

Although in use since the 1990s, the term big data was launched into the public sphere (i.e. became viral) at nearly the same time as the terms data science and data scientist: around 2008, when the British weekly scientific journal *Nature* published a special issue entitled "Big Data: Science in the Petabyte Era" on the tenth anniversary of Google's incorporation ("Community Cleverness Required" 2008). By this time, Google's enormous success as a company founded on data mining had caught the world's attention, including that of the scientific community, to the point where the company had become something of a paradigm for science. The issue was devoted to exploring how science ought to manage and exploit big data by following Google's lead through various data processing methods, from data mining to visualization to library science. This connection between Google and science was also made by *WIRED*'s Chris Anderson at this time, in an issue also devoted to "the Petabyte Age," who argued:

Our ability to capture, warehouse, and understand massive amounts of data is changing science, medicine, business, and technology. As our collection of facts and figures grows, so will the opportunity to find answers to fundamental questions. Because in the era of big data, more isn't just more. More is different (Anderson 2008b).

Anderson argued that Google's successful application of model-free algorithms, as in its ad auctioning system, showed that the scientific method was obsolete; or, more accurately, that science might "learn from Google" and by-pass the concern for theory building and focus on prediction. The parallel to Breiman's characterization of the algorithmic modeling culture is clear here.

The rise of the term big data is indicative of an important shift in how the problem of data impedance was conceptualized. Since at least the 1960s, when the trope "data deluge" was invented apparently by NASA, the problem of data surplus was always framed as a kind of disaster, as is evident from the image of a flood, and the semantically close and more popular phrase "information explosion," implicitly likened to nuclear weapons by the frequent use of the image of the mushroom to signify exponential growth. With the success of the data-driven corporation on the model of Google, these negative terms began to be replaced by the more positive, or at least neutral, expression big data. In fact, one may observe this transition in the simultaneous publication of the *Nature* and *WIRED* issues on the topic (cited above)—the former introduces the new term while the latter uses the old, and both are linked by the metonym of the "petabyte" era or age. Since then, the term big data has been used to signify the context and opportunity within which the data science operates. For example, Patel and Davenport 2014 article defined a data scientist as "a high-ranking professional with the training and curiosity to make discoveries in the world of big data." This connection became a commonplace. In 2013, *Communications of the ACM* published "Data Science and Prediction," which also directly linked the big data to data science, while providing some flesh to the former:

> … data science is different from statistics and other existing disciplines in several important ways. To start, the raw material, the "data" part of data science, is increasingly heterogeneous and unstructured text, images, video often emanating from networks with complex relationships between their entities. ... Analysis, including the combination of the two types of data, requires integration, interpretation, and sense making that is increasingly derived through tools from computer science, linguistics, econometrics, sociology, and other disciplines. The proliferation of markup languages and tags is designed to let computers interpret data automatically, making them active agents in the process of decision making. Unlike early markup languages (such as HTML) that emphasized the display of information for human consumption, most data generated by humans and computers today is for consumption by computers; that is, computers increasingly do background work for each other and make decisions automatically. This scalability in decision making has become possible because of big data that serves as the raw material for the creation of new knowledge; Watson, IBM's "Jeopardy!" champion, is a prime illustration of an emerging machine intelligence fueled by data and state-of-the-art analytics.

Here, big data is linked to both data science and to the *kinds* of data that have been associated with the field since the AFCRL, in addition to textual data specific to the Internet and the Web.

### 3.7. 2012: The Disconnect with Statistics

Among the most significant developments in the years immediately following the emergence of what I have called big data science was the explicit perception by professional statisticians that all of this occurred independently of their field, and that statisticians would do well to take advantage of the new interest in data that was sweeping the business world. In a series of surprisingly candid editorials in *AmStatNews*—the membership magazine of the American Statistical Society—no fewer than three succeeding presidents of the organization, from 2012 to 2014, offered their views on what they saw as a troubling "disconnect" between the field of statistics and data science.

This disconnect—between the self-perception among statisticians that they already are data scientists and their exclusion from real developments in industry and the media under the name of big data—is captured by this anecdote given by Marie Davidian in her column (entitled "Aren't We Data Scientists?"):

> I was astonished to review the list of founding members [of the National Consortium for Data Science (NCDS) based in North Carolina] and see that not only is my university (North Carolina State) a founding member, but so are Duke University and UNC-CH. Along with SAS Institute; Research Triangle Institute International; NIH's National Institute for Environmental Health Sciences; IBM; and several other institutions, businesses, and government agencies that employ numerous statisticians. The member representatives listed on the website from NC State, Duke, and UNC-CH are computer scientists/engineers, and among all 17 representatives, *there is not one statistician*. (Davidian 2013: 3; emphasis added.)

The gap was noted a year earlier by Robert Rodriguez, but without the surprise:

> A recurring theme in Big Data stories is the scarcity of "data scientists"—the term used for people who can draw insights from large quantities of data. This shortage was highlighted in an April 26, 2012,

*Wall Street Journal* article titled, "Big Data's Big Problem: Little Talent" (Rooney 2012). The question "What is a data scientist?" is still being debated (see the articles with this title at Forbes). However, *there is consensus that data scientists must be innovative problem solvers with expertise in statistical modeling and machine learning, specialized programming skills, and a solid grasp of the problem domain*. Hilary Mason, chief data scientist at bitly, adds that "*data scientists are responsible for effectively communicating the things that they learn. That might be creating visualizations or telling the story of the question, the answer, and the context*." (Rodriguez 2012: 3-4; citation and emphases added.)

It is notable that Rodriguez clearly recognized the reality behind the disconnect, conceding that "our profession and the ASA have not been very involved in Big Data activities." He did not trivialize the concepts of big data and data science; instead, he patiently explained their distinctive features and provided suggestions for how statisticians can add value to these developments going forward. He suggested that statisticians should "view data science as a blend of statistical, mathematical, and computational sciences," and focus their efforts on how to "extract value from data not only by learning from it, but also by understanding its limitations and improving its quality. Better data matters because simply having Big Data does not guarantee reliable answers for Big Questions."

In a subsequent editorial co-authored with the two succeding presidents of the ASA, Rodriguez's recognition of the absence of statistics from data science and his strategy to focus on what statisticians do best is amplified and augmented:

> Ideally, statistics and statisticians should be the leaders of the Big Data and data science movement. Realistically, we must take a different view. While our discipline is certainly central to any data analysis context, *the scope of Big Data and data science goes far beyond our traditional activities.* As Bob [Rodriguez] noted in his column, the sheer scale and velocity of the data being generated from multiple sources requires new data management and computational paradigms. New techniques for analysis and visualization must be developed. And communication and leadership skills are critical.

> We believe we should focus on what we need to do as a profession and as individuals to become valued contributors whose unique skills and expertise make us essential members of the Big Data team. . . . We know statistical thinking—our understanding of modeling, bias, confounding, false discovery, uncertainty, sampling, and design—brings much to the table. We also must be prepared to understand other ways of thinking that are critical in the Age of Big Data and to integrate these with our own expertise and knowledge.

> We have had many discussions—among ourselves and with ASA members who are familiar with Big Data—about strategies for achieving this preparation and integration. These discussions have led to our joint ASA presidential initiative to establish the statistical profession as a valued partner in Big Data activities and to position the ASA in a proactive and facilitating role. *The goal is to prepare members of our profession to collaborate on Big Data problems.* Ultimately, this preparation will bridge the disconnect between statistics and data science. (Rodriguez, Davidian, and Schenker 2013)

Not all academic statisticians were willing to concede the point that data science "goes far beyond our traditional activities." Indeed, many viewed data science as an invader of their territory. Bin Yu, then president of the Institute of Mathematical Statistics, exhorted her colleagues to "own data science" (Yu 2014), echoing Davidian's exasperatated observation (and no doubt the sentiment of many) that statisticians "already are" data scientists. To make her point, she defined the core components of data science—statistics, domain knowledge, computing, teamwork, and communication—and then traced each of these to the traits of various ancestors in her field. In this narrative, Harry Carver, Herman Hollerith, and John Tukey are all data scientists *avante la lettre*. Indeed, Carver emerges as an "early machine learner," which allows Yu to consider machine learning as an element of statistics.

### 3.8. The Academic Response

After 2012, the field of data science and the cluster of associated activities associated with it grew exponentially. As mentioned above, this growth was associated with a high demand for data scientists, a story that continues to be covered by the news media. The response by institutions of higher education to train data scientists to meet industry demand was rapid and pronounced. Hundreds of master's degree programs in data science and closely related fields were established in the United States, often associated with the formation of institutes of data science. More recently, a handful of doctoral programs and schools of data science have emerged, along with undergraduate offerings to meet increasing student demand. The trend to create degree programs for the field continues.

One effect of these developments has been to stimulate a preferential attachment process within the network of disciplines that constitute the academy: as a field representing the "sexiest job of the 21st century," attracting students, gifts, and internal resources, many adjacent disciplines—from systems engineering and computer science to statistics and a variety of quantitative and computational sciences—have sought to associate themselves with the field. Indeed, because data science per se has had no history in the academy, these contiguous fields have provided the courses and faculty out of which the majority of data science programs have been built. The result is that data science has become a complex and internally competitive patchwork of industrial and academic interests and perspectives, reflecting the broader engagement of society with data and its analysis beyond the concept of data science inherited from industry.

Yu's argument that statistics should take over data science is made later, and more thoroughly, by Donoho in "50 Years of Data Science" (Donoho 2017), which is essentially a manifesto for the annexation of data science by statistics in response to the proliferation of academic programs associated with the new field. He charts out the territory of "Greater Data Science" (GDS)—a play on Chambers' earlier plea for a "greater statistics" that would be "based on an inclusive concept of learning from data"—and places statistics at its center (Chambers 1993: 1). He locates GDS in a genealogy that begins with data analysis—a practice envisioned in the 1960s by his mentor at Princeton, the legendary mathematician John Tukey, who serves as the founding ancestor in this legitimation narrative. GDS is thus defined as "a superset of the fields of statistics and machine learning, which adds some technology for 'scaling up' to 'big data'" (Donoho 2017 [2015]: 745). He also attempts to deflate the concept of big data, so central to contemporary data science, by citing the  Holerith's punched card system, which was invented to process the unexpectedly large volume of data produced by the 1890 US census, as an early instance and therefore nothing new. Like Yu, Donoho's argument is that, beyond the introducing a few useful technologies, data science as a whole is nothing new. It is a scandal that it has emerged outside of the field of statistics and is represented in the media as a distinct field.

Donoho's essay has been criticized for downplaying the contribution of computational technology to data science. In his response to the essay, Chris Wiggins, Chief Data Scientist of the *New York Times* and a professor at Columbia, sensed this and asserted that data science is a form of engineering that will be defined by its practitioners, not by academics trying to turn it into a (pure) science. In his response, Sean Owen, Director of Data Science at Cloudera, argued that Donoho's history excluded the significant contributions of data engineering (Donoho 2017: 764). Elsewhere, Bryan and Wickham pointed out that, like many statisticians, Donoho mistakenly relegated computational work to superficial status while also missing "the full process of analysis" in which statistics "is but one small part" (Bryan and Wickham 2017). In his defense, Donoho acknowledged his bias, but justified it by noting that although technological know-how is important, technologies are transitory and prone to rapid obsolescence, and therefore "the academic community can best focus on teaching broad principles—'science' rather than 'engineering'" (Donoho 2017: 765). *Scientia longa, brevis ars.*

It is beyond the scope of this essay to fully evaluate the arguments of Donoho and Yu. Suffice it to say that when Facebook was hiring data scientists in 2008, the fact that someone's academic field could claim Hollerith and Carver as data scientists would not have improved that candidate's chances of being hired. What is significant here for understanding the history of data science is the social underpinning of the observed disconnect between members of the established field of statistics and those of the emerging one of data science. By 2014, Breiman's observation that the developments in algorithmic modeling, and more generally data mining, "occurred largely outside statistics in a new community," was proven to be both true and prescient: that new community became one of the primary tributaries to data science, and the long standing opposition of statistics toward the beliefs of that community—evident in Donoho and the predecessors he cites in the 1990s and early 2000s—became manifest.

## 4. Interpretation

Let us stop here, roughly at the point where data science becomes widely known and legitimate in the eyes of industry, and both accepted and contested within academia, and which characterizes the current period. What can we glean from this historical outline? Several themes emerge. Most significantly, it has been established that the term data science per se dates back to the early 1960s with the formation of the Data Sciences Laboratory in Cambridge, Massachusetts, and this usage was surprisingly close to its current one. To review, the esseential elements were there: the presence of big data (properly understood) and the use of both computational machinery and artificial intelligence to make sense of it. Morevoer, this original meaning remains surprisingly consistent in the decades that follow, even as the term developed and accreted new senses. This development can be characterized as having three main phases resulting in three major variant definitions of the term: (1) Computational Data Science, (2) Statistical Data Science, and (3) Big Data Science. In addition, we can add a fourth definition, inchoate and currently being developed, that we may call Academic Data Science. These are described below.

### 4.1. Four Definitions

#### 4.1.1. DS₁: Computational Data Science

This is the original, or classical, definition of data science that begins with the Data Sciences Laboratory and is taken up, at first in spirit if not in name, by organizations like CODATA during the same era. Data science in this definition is *the science of data in support of science and decision-making*. This definition also includes the concept of datalogy developed by Naur as well as the data processing know-how of corporations like Mohawk Data Science, but it is primarily a field developed by and for scientists and engineers to handle the problems arising from data surplus, culminating in the so-called fourth paradigm of science. That this field by this name persists and becomes established through the current era is evident in the fact that the term "data scientist" had currency in places like *New Scientist, TheTimes* of London, and other news sources in the 1990s and 2000s. In addition, data science in this sense is the subject of high level reports from the NSF and JISC in the 2000s. That this definition persists to this day can be found in examples like the work of the Dutch data scientist Demchenko, who assigns data management, curation, preservation central places in the definition of data science (Demchenko 2017).

Essential to this definition is a focus on what is eventually called big data and issues arising from its curation and processing. Importantly, this definition also frames data surplus as a positive condition that makes possible new ways doing science, i.e. the fourth paradigm view of science. Methodologically, this definition also embraces AI, machine learning, and data mining methods that may or may not be principled from a strict statistical point of view. It also embraces statistical methods, but from the practical perspective of scientists are who often not concerned with the concerns of pure, mathematical statisticians.

Arguably this definition also includes adjacent work in computer science of data processing, information retrieval, and information science which led to the invention and development of databases and general theories of data. Eventually, the fruits of this work would lead to the conditions of data impedance that led to data mining, a practice that converted the vice of surplus into a virtue by establishing a mutually beneficial relationship between surplus digital data and machine learning.

### 4.1.2. DS₂: Statistical Data Science

By statistical data science, I refer to the usage developed by Hayashi and Ohsumi (the Tokyo School) and the American statisticians Wu, Kettenring, and Cleveland, as well as Donoho. These statisticians implored their colleagues, unsuccessfully, to adopt the term data science in order to rebrand their discipline in response to the overshadowing effects of computational statistics and data mining that were being felt in the 1990s.

The essential characteristics of this definition are the renewed commitment to Tukey's conception of data analysis and, more generally, an appreciation of the foundational role of data in statistics, along with an exhortation to take seriously recent developments in computer science in areas ranging from machine learning to databases. However, these technologies are to be incorporated with the admonition to avoid the practices of data mining, which, on its own, is considered unprincipled. Indeed, this definition may be seen primarily as an effort to correct what are perceived to be the fruitful but misguided efforts of data mining by grounding its computational methods in a mathematically sound framework.[20]

Regarding the relationship of data science to the computational technologies on which it depends, this definition treats them as essential but external to core practice. Languages, servers, and databases are thought of as an environment within which the analyst carries out an essentially mathematical set of tasks with greater efficiency, not as the medium through which one thinks about the world. Their net effect on the work of statistics is considered to be a difference in degree, not in kind.

### 4.1.3. DS₃: Big Data Science

By big data science, I refer to the form of data science that emerged in the context of web companies like Google and Facebook and become both viral and paradigmatic after being anointed by *HBR* in 2012. As we have seen, this definition inflects the classical, computational definition in the context of the Silicon Valley social media firm, or what Zuboff has called surveillance capitalism. The conditions of data impedance that attended the rise of Big Science after WWII are embraced and become the foundation of a new business model that in turn becomes a model for all other firms and sectors to imitate, from the automotive industry to medicine.

---

[20] The works of Hastie, Tibshirini, and their coauthors are perhaps the most successful exemplars of this definition; their work incorporates enthusiastically data mining, but always within the encompassing framework of statistical thinking that effectively domesticates the field (Hastie, Tibshirani, and Friedman 2009; Efron and Hastie 2016).

One of the distinctive features of this definition is the close association with big data, in perception if not always in reality, as both a set of new technologies to manage 3D data and a set of "unreasonably effective" methods to convert these data into value. Another feature is the focus on data wrangling, the work required to convert the widely varying formats and conditions of data in the wild into the standard analytical form. In addition to these features, and consistent with Varian's remarks in the 2008 McKinsey interview, big data science embraces a suite of activities that connect these practices to the context of business and decision-making, such as visualization, communication, business acumen, and a focus on marketable data products.

It follows that Donoho is incorrect in asserting that "[w]e can immediately reject 'big data' as a criterion for meaningful distinction between statistics and data science" (Donoho 2017: 747). The assemblage of computational technology associated with big data is the condition of possibility of data science in this definition, its *sine qua non*. It is impossible to imagine this kind of data science without the infrastructure of data generating machinery, high-performance computing architectures, scalable database technologies such as Hadoop and its descendants, and data-savvy programming languages such as R, Python, and Julia, and, to a high degree, the availability of extant datasets on the web. Indeed, the connection between this kind of data science and the technology stack on which it stands is so close that the relationship between technology and science becomes blurred, leading to revolutionary proclamantions of new kinds of science and conservative reactions to such claims (such as Donoho's).

### 4.1.4. DS₄: Academic Data Science

By academic data science, I refer to the ongoing reception of big data science by the academic community in response to the demand for data scientists across all sectors of society. The idea of data science has influenced the academy in two ways: first, by stimulating the production of degree programs to meet workforce demands, and second, by providing a model for effective knowledge production in the context of pervasive data and data technologies. Both of these influences have given rise to the secondary effect of bringing to the surface and aggregating the myriad other forms of data-centric and analytic activities already being conducted in the academy (and elsewhere) for years, from statistics to operations research to e-sceince, all of which have claims to be "already doing data science." This situation has led to current crisis in the definition of data science that is has produced responses such as Donoho's as well as the current essay.

Based on personal experience, I believe that the term data science within the academy has in recent years been nudged in the direction of being identified with an expanded form of statistics (DS₂), regardless of whether its programs are "owned" by departments of statistics or not. This is because of the great authority of the field of statistics within the academy, as well as a general skepticism toward industry-generated categories by academics, many of whom dismiss the terms big data and data science as buzzwords. This has put pressure on developers of data science programs to become academically legitimate in the eyes of their peers and administrators. The shift in meaning is also due, quite frankly, to the co-opting of the term by departments of statistics to both cash in on the term and stem the tide of what are perceived to be its negative qualities. This tendency has had the effect of flooding the market of data scientstis with de facto data analysts who are unable to perform the work that many in industry had previously sought under the sign of data scientist. And this has produced a counter-effect within industry to invent a new category—the data engineer. In reality, this category is surprisingly close in meaning to the original category of data-processing scientist that was coined in the 1950s in the context of data reduction and other work that eventually became associated with the AFCRL Data Sciences Laboratory and with scientific research data management in general.

This is an unfortunate state of affairs. The great value of data science has been in its cultivation of the fertile land between the science and engineering of data opened up by the great advances in data generating and processing machinery for both science and industry. It is also unfortunate because as the academy produces a definition of data science at odds with what science and industry need, the latter are once again left to fend for themselves, creating their own ad hoc educational resources to convert computer scientists and other adjacent role into data engineers. A rose is a rose by another other name.

## 4.2. Data Impedance

To summarize, the field of data science has a surprisingly consistent and durable history, even if, on the ground, the individual actors in this social drama have not been not aware of this fact. The original constitution of the field, $DS_1$, survives to the present day, providing the backbone and the foil for each of the following configurations. For if $DS_2$ is clearly a reaction to the effects of the first, $DS_3$ is a revival of $DS_1$ in a new key. Wu and Hammberbacher may not have been aware that they were borrowing a term from one context and applying it to another analogous one, but social facts are rarely perceived as such by the individuals who participate in them.

I hypothesize that the source of this continuity is a persistent situation in which it makes sense to use the term data science. The term is motivated in at least two ways. First, it is motivated semiotically by virtue of its complementary relationship to other extant categories which constitute the repertoire available concepts and terms. Data science is not computer science nor information science nor statistics nor data analysis but adjacent to each of these. In each case, the term was selected from the sample space of these other terms, which always remained possible choices. The fact that they were not selected is what is significant. Second, the term was and is motivated by the role that it plays in referencing a persistent assemblage of material conditions that emerged during the post war era and continues to this day. In addition to being embodied by the SAGE system that arguably motived the initial coinage of data science, those material conditions gave birth to the Internet, whose construction was first conceived shortly after 1957 in response to the launch of Sputnik by the Soviet Union, and to the field of data processing and everything associated with it, including the development of databases and the refining of the concept of data itself.

I have chosen the term data impedance to characterize this persistent situation. Again, by this term I mean the disproportion between the surplus of data generated by the machinery of data production, and the relative scarcity of computational power and methods to process these data and extract value from them. To be sure the condition of impedance has been part of the human condition since the formation of states which require the use of media to function. By media, I mean external records that must be stored and interpreted to be useful. Such records range from the quipus of the ancient Inca, to the hieroglyphic writing systems of the ancient Maya and Egyptians, to those of Asia and Europe. What is historically unique about the post-war condition of data impedance is that it occurs within the milieux of digital and electronic data characterized as the input and output of computational machinery. Although other forms of data are clearly part of the condition I am describing, these technologies are at the center and what gives the unique historical character to the condition I am describing and to data science.

## 4.3. Data mining vs data analysis

A final theme one may observe in this history is that data science has been a contested term at least since the 1990s, when statistical data science ($DS_2$) emerged in Japan and the US in response to the developments of classical, computational data science. As we have seen, this response was partly an attempt to embrace the advances made by computational data science ($DS_1$) and partly an effort to correct what were perceived to be the excesses of this approach. The marginalization of computational technology in this definition of data science is consistent with the larger conflict between what Breiman famously called "two cultures" in the work of statistics, adapting the expression C.P. Snow used to characterize the split between the sciences and the humanities (Breiman 2001; Snow 2013 [1959]).[21]

Breiman's trope provides a useful framework for capturing the epistemological differences between the two communities associated with these definitions. On the one hand, we have the data analysts, on the other the data miners. The former, descendants of Tukey, remained faithful to the mission of statistics to provide a mathematically principled methodology for working with data. Ideally, all data were understood to be produced by information generating mechanisms that can be described by interpretable models and, in the ideal case, parametrically. Even Bayesian methods, long held back by their complexity, but reborn with the rise of computational methods like MCMC and Gibbs Sampling, were reined in by the data modeler's ethos. The latter, the data miners, unfettered by such requirements, enthusiastically applied the newer and rapidly developing world of algorithms and, more generally, computational thinking to the data surplus that was inundating science, government, and industry.

The differences between these two groups may be characterized in many ways, such as how they conceive of the relationship between data and motivating questions. Another way, implied by the preceding account, is in their relation to computing technology. For the data analyst, the computer is an external *environment* that supports their work; the computer is convenient but not necessary, at least in principle, as Tukey implied by his reference to "pen, paper, and slide rule." For the data miner, the computer is an immersive *medium* within which their work is made possible; it is a *sine qua non* for their work.

The effects of media on thinking have long been recognized and studied, especially in relation to differences between orality and literacy. In this case, one effect of the medium of data may be noted: for the computational thinker, the separation between data and models is always provisional. This is because models may always be represented as code and are therefore as data, given the equivalency between the two in the Von Neuman architecture, the common ancestor of practically all of computer processor technology in use today (Dyson 2012). This equivalency makes homoiconic languages, such as LISP not only possible, but natural. It also predisposes the computational thinker, of which the data miner is a special case, to apply the same modeling procedures to code that are applied to data elsewhere. Here we are far beyond the treatment of parameters as subject to probability, as Bayes' theorem requires; every aspect of a statistical model may be subject to combinatorial play. This helps to explain the genesis of many data mining approaches, drawing from artificial intelligence, such as genetic algorithms, where model formulae themselves are manipulated as long strings.

---

[21] It is no coincidence that Breiman's essay appeared at about the time some academic statisticians sought to rebrand their field as data science in an attempt to integrate the gains of computational technology while purging it of the methodological sins of data mining. Although Breiman is sometimes counted among those wishing to rebrand statistics as data science, the substance of his remarks, which did not reject the spirit of data mining, went against the grain of that movement.

To conclude, an authentic definition of data science would embrace the term's history. As we have seen, this history is not merely etymological; the term indexes a persistent situation that continues to motivate the current practice of data science in industry and science. Academics would do well to embrace this and avoid what may be called the fallacy of purism as we seek to make sense of the field as a body of knowledge. This means embracing the oppositions between data analysis and data mining, and between science and engineering, as a core, animating tensions in the field that may be cultivated for its generativity.